%% file: report.tex
\documentclass[11pt]{article}
\usepackage{graphicx} 
\usepackage{defs,fonts}
\usepackage[margin=1in]{geometry} 
\usepackage{algpseudocode}
\usepackage{afterpage}
\usepackage{amsbsy}
\usepackage{amsmath}
\usepackage{amssymb}
\usepackage{amsthm}
\usepackage{bm}
\usepackage{bbm}
\usepackage{caption}
\usepackage{cases}
\usepackage{cite}
\usepackage{color}
\usepackage{dcolumn}
\usepackage{dsfont }
\usepackage{fonts}
\usepackage{graphicx} 
\usepackage{microtype}
\usepackage{subcaption}
\usepackage{tabularx}
\usepackage{tikz} 
\usepackage{tikz-3dplot}
\tikzset{>=latex}
\usetikzlibrary{calc}
\usepackage{cleveref}
\usepackage{booktabs}       
\usepackage[ruled,vlined,linesnumbered,noend]{algorithm2e}
\usepackage{amsfonts}
\usepackage{comment}
\usepackage{tikz}
\usepackage{tikz-3dplot}

\usepackage{wrapfig}
\usepackage{booktabs}    
\usepackage{adjustbox}
\usepackage{graphicx}
\usepackage{array}
\usepackage{url} 

\title{Reference solutions for linear radiation transport: \\ the Hohlraum and Lattice benchmarks\thanks{
This material is based upon work supported by the U.S. Department of Energy, Office of Science, Office of Advanced Scientific Computing Research, as part of their Applied Mathematics Research Program. The work was performed at the Oak Ridge National Laboratory, which is managed by UT-Battelle, LLC under Contract No. De-AC05-00OR22725. The United States Government retains and the publisher, by accepting the article for publication, acknowledges that the United States Government retains a non-exclusive, paid-up, irrevocable, world-wide license to publish or reproduce the published form of this manuscript, or allow others to do so, for the United States Government purposes. The Department of Energy will provide public access to these results of federally sponsored research in accordance with the DOE Public Access Plan (http:/energy.gov/downloads/doe-public-access-plan).
This research used resources of the Compute and Data Environment for Science (CADES) at the Oak Ridge National Laboratory, which is supported by the Office of Science of the U.S. Department of Energy under Contract No. DE-AC05-00OR22725"
}}

\author{Steffen Schotth\"ofer\footnote{Corresponding author. Mail to \texttt{schotthofers@ornl.gov}}\, and Cory Hauck \\
\,\\
Computer Science and Mathematics Division \\
  Oak Ridge National Laboratory \\
  Oak Ridge, TN 37831 USA}
\date{\today}

\begin{document}
\input{macros_steffen}

\maketitle

\begin{abstract}
    Two benchmark problems for linear radiation transport and derived from the literature are presented in detail and several quantities of interest are defined. High-resolution simulations are computed using standard, robust numerical methods  and implemented using HPC resources.   The goal of these simulations is to provide reference solutions for new discretization approaches, to aid in the development of multi-fidelity uncertainty quantification and optimization algorithms, to provide data for training surrogates models.  The code used to perform the simulations and post-process the data is publicly available, as is the raw data that used to generate the results presented in the paper.
\end{abstract}


\section{Introduction}
Radiation transport describes to the propagation of energetic particles through space as they interact with a surrounding material medium.  The study of radiation transport has applications in a wide variety of fields, including medical imaging \cite{suetens2017fundamentals}, cancer therapies \cite{baskar2012cancer}, and astrophysical flows \cite{mihalis2013foundations}. 
Radiation transport is also key to understanding the movement and transfer of energy in engineering devices, for example by neutrons in fission reactors \cite{davison1957neutron,lewis2008fundamentals} or by photons that are used to heat inertial confinement fusion capsules \cite{hurricane2023physics}.  

In a kinetic description, radiation transport is modeled by a radiation transport equation (RTE) that prescribes the density of the radiation in position-momentum phase space.   
The challenges involved with solving RTEs numerically are well known.  
First, an RTE solution may be a function of (as many as) six phase-space dimensions (three space and three momentum) and may also depend on time. 
Thus standard, grid-based discretization methods require a great deal of memory and a large number of floating point operations to obtain an accurate solution, while Monte-Carlo approaches are known to suffer from noise \cite{dupree2012monte,lewis1984computational}.  
Second, the interactions of radiation with the material can be extremely complex, especially if the material is also evolving \cite{mihalas1984radiation}.  Third, variations in collision rates, due to multiple materials or strong dependencies on energy, mean that the RTE can support multiple spatial and temporal scales \cite{bensoussan1979boundary, larsen2023asymptotic}.  Resolving all of these scales increases the computing requirements even further.

A variety of numerical tools have been developed to reduce the computational expense of simulating RTEs.  
These methods continue to make high-fidelity simulations more tractable, especially when combined with modern computing resources. 
Some of the major advances include (i) asymptotic preserving methods \cite{ larsen1989asymptotic,larsen1987asymptotic,jin2010asymptotic} that can compute accurate solutions to collision-dominated flows with under-resolved meshes, (ii) sophisticated iterative schemes that enable implicit time integration with large time steps \cite{warsa2004krylov,adams2002fast,chacon2017multiscale,gol1964quasi}, and  (iii) low-memory approaches \cite{yin2025towards, peng2022reduced, peng2020low,widmer2008sparse,grella2011sparse} that reduce the number of unknowns needed to discretize the phase space.

The purpose of the current paper is to provide quantities of interest, computed from highly resolved solutions to two benchmark problems.  These two benchmarks, both derived from \cite{brunner2002forms}, do not possess exact solutions; moreover, the construction of a manufactured solution may require a non-physical source that is not desirable, especially if it spoils the physical nature of the solution.  Thus the goal of this computational study is to provide a highly resolved reference solution for testing newer, more cost efficient methods that are currently being developed in the research community. Such methods are needed for ``outer loop" computations, such as uncertainty quantification, design, and control -- tasks which require many forward simulations, but cannot afford the level of resolution provided here.  
We construct high resolution solutions for both benchmarks using the highly parallel  \texttt{C++}-based solver KiT-RT~\cite{kusch2023kit} using a discrete ordinate ($S_N$) discretization in angular space, a finite volume method on an unstructured for physical space, and an explicit SSP Runge-Kutta time stepping scheme.  These numerical tools are themselves not novel; however, with sufficient resolution, they provide stable and robust numerical solutions with minimal artifacts.    

The first benchmark problem, referred to as the Lattice benchmark, is a toy mock-up of a nuclear reactor assembly; it is taken directly from \cite{brunner2002forms}.  The second problem is a linearized, symmetric Hohlraum benchmark is a variation on the original nonlinear hohlraum problem first presented in \cite{brunner2002forms}.  The linearized version of this problem was first introduced in \cite{osti_1550352} in order to allow computational studies without directly simulating the evolution of the material temperature, which in turn affects scattering and absorption rates.  In the current work, we further modify the linearized problem to incorporate a symmetric geometry that is more like the double-ended hohlraum design from \cite{brunner2007estimating}. 

For both benchmarks, the RTE is linear, and all particles belong to a single energy group with unit speed.  Moreover, the geometry is two-dimensional and includes only rectangular features, which further reduces the size of the phase-space problem and simplifies meshing requirements.  In particular, the phase space is only four-dimensional.  Even so, the computational resources required are still substantial.

The remainder of this paper is organized as follows.
In \Cref{sec_rad_transport}, we present the radiation transport equation. In \Cref{sec_problems},  we formulate the two benchmark problems:  Lattice and Hohlraum. In \Cref{sec_numerics}, we present the discretization scheme that is used to compute numerical solutions for these problems. In \Cref{sec_implementation}, we describe implementation and solver settings in details. In \Cref{sec_results}, we present the simulation results.   In \Cref{sec_conc}, we make conclusions and provide discussion.

\section{The radiation transport equation}\label{sec_rad_transport}

In this section, we present the radiation transport equation (RTE).  We then introduce the symmetry assumptions that allow the physical domain of the RTE to be reduced to two spatial dimensions.

\subsection{General formulation}
Let $X\subset \mathbb{R}^3$ be a material domain with Lipschitz boundary $\p X$ and let $\bbS^2$ be the unit sphere in $\bbR^3$.  Let 
\begin{align}\label{eq_3d_boundary}
    \Gamma^{\pm} = \{ (\bx,\bsOmega) \in \p X \times \bbS^2 \colon \pm \bn(\bx) \cdot \bsOmega > 0 \}
\end{align}
be the inflow (-) and outflow boundaries (+), where $\bn(\bx) \in  \bbR^3$ is the outward unit normal vector to $X$, defined for almost every $\bx \in \p X$.
Let $\Psi(\bx,\bsOmega,t)$ be the density of particles --- with respect to the phase space measure $d \bsOmega d \bx$ --- at position $\bx\in X$ and traveling in the direction  $\bsOmega$ at time $t \geq 0$. 

We consider the case of particles with unit speed that interact with a static material via scattering and absorption, at rates characterized by cross-sections $\sigmas=\sigmas(\bx)$ and $\sigmaa = \sigmaa(\bx)$, respectively.  In particular the scattering is assumed to be isotropic, i.e., independent of $\bsOmega$. The material also generates a known isotropic particle source $Q=Q(\bx)$.   Under these assumptions, $\Psi$ is governed by an RTE of the form
\begin{align}\label{eq_transport}
\begin{cases}
  	\p_t \Psi+\bsOmega \cdot \nabla_{\bx}\Psi +\sigmat \Psi 
        =  \frac{\sigmas}{4\pi}\vint{\Psi} +Q,  
        &\quad (\bx , \bsOmega) \in X \times {\mathbb{S}^2}, \;  t >0,  
        \\
    \Psi(\bx,\bsOmega,t) =  \Psi^-(\bx,\bsOmega,t) 
    & \quad   (\bx , \bsOmega) \in \Gamma^- , \; t>0, \\
    \Psi(\bx,\bsOmega,0) =  \Psi_0(\bx,\bsOmega) 
    & \quad (\bx , \bsOmega) \in X \times {\mathbb{S}^2},
 \end{cases}
\end{align}
where $\vint{\Psi} = \int_{\bbS^2} \Psi d \bsOmega$ and $\sigmat = \sigmas + \sigmaa$.  The initial condition $\Psi_0(\bx,\bsOmega)$ and  inflow boundary data $\Psi^{-}  = \Psi^-(\bx,\bsOmega,t)$ are given.

\subsection{Reduction to two dimensions}\label{sec_2d_redux}
\begin{figure}[h!]   
\centering
\begin{subfigure}[b]{0.45\textwidth}
\centering
\tdplotsetmaincoords{60}{110}
\begin{tikzpicture}[scale=5, tdplot_main_coords]
    \coordinate (O) at (0,0,0);
    \draw[thick,->] (0,0,0) -- (1.1,0,0) node[anchor=north east]{$\be_x$};
    \draw[thick,->] (0,0,0) -- (0,0.8,0) node[anchor=north west]{$\be_y$};
    \draw[thick,->] (0,0,0) -- (0,0,1) node[anchor=south]{$\be_z$};
    \tdplotsetcoord{P}{1}{30}{60}
    \draw plot [mark=*, mark size=0.2] (P) node [right] {\scriptsize$(\theta,\phi)$};
    \draw[->, thick] (O) -- (P) node [midway, below right] {$\bsOmega$};
    \draw[dashed, color=black] (O) -- (Pxy);
    \draw[dashed, color=black] (P) -- (Pxy);
    \tdplotdrawarc{(O)}{0.2}{0}{60}{anchor=north}{$\varphi$}
    \tdplotsetthetaplanecoords{60}
    \tdplotdrawarc[tdplot_rotated_coords]{(0,0,0)}{0.4}{0}%
        {30}{anchor=south}{$\theta$}
 \end{tikzpicture}  
\caption{Spherical coordinates}
\label{fig:spherical-coordinate}
\end{subfigure}%
\hfill
\begin{subfigure}[b]{0.45\textwidth}
\centering
\includegraphics[width=0.8\textwidth]{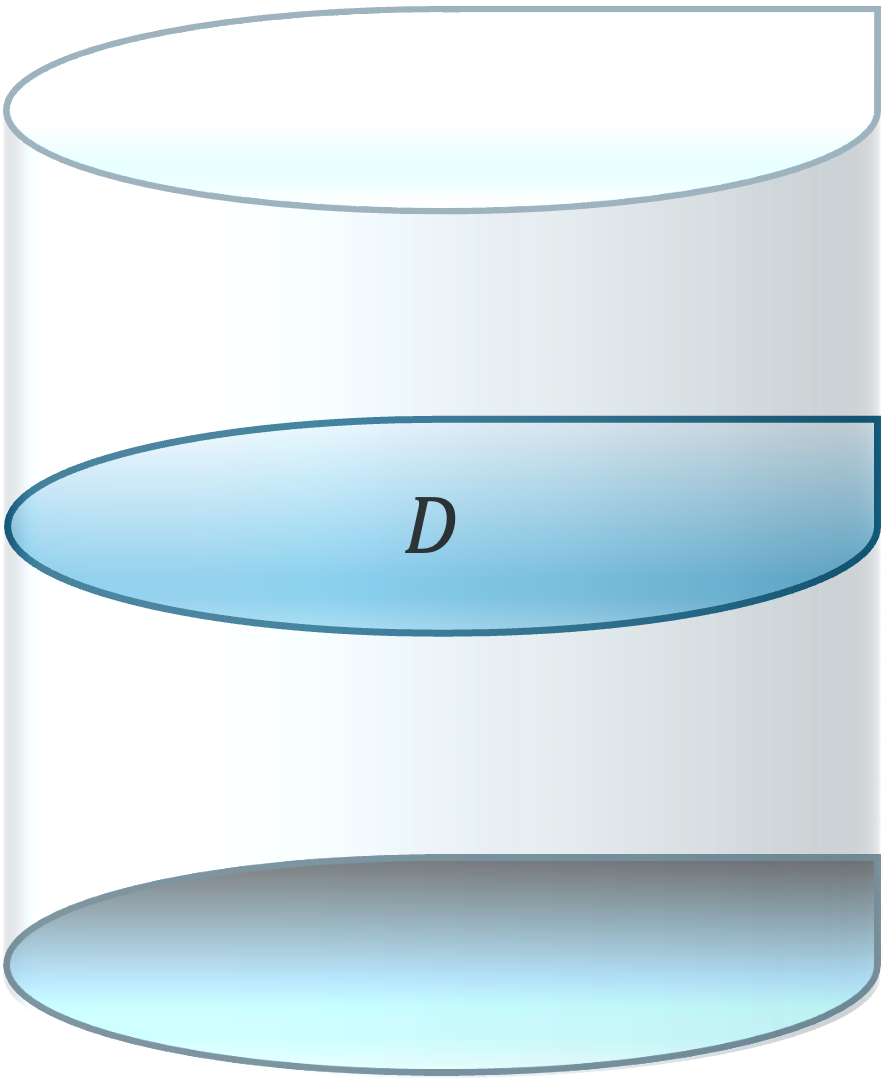}
\caption{$X = D \times \bbR$.}
\label{fig:D}
\end{subfigure}
    \caption{Geometry for the reduction to two space dimensions.}
    \label{fig:geometry}
\end{figure}
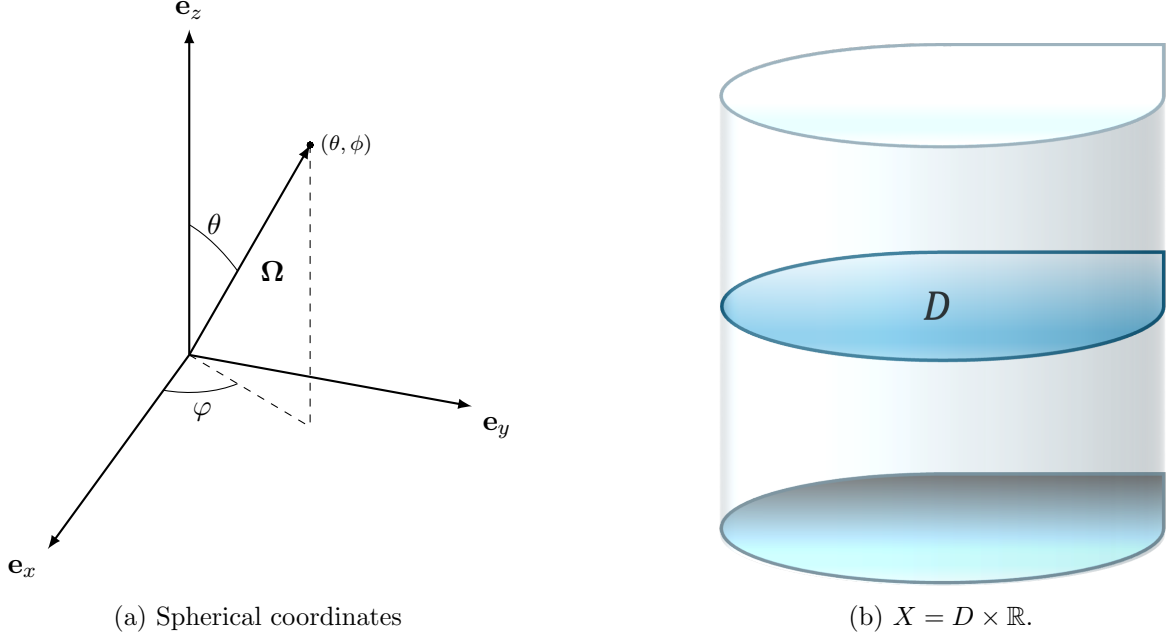

The benchmark problems in this paper assume symmetries which allow the RTE to be recast as an equation in only two spatial dimensions.  Let $\be_x$, $\be_y$, and $\be_z$ be the canonical Cartesian unit vectors.  Write $\bx = x \be_x + y\be_y + z \be_z$ and $\bsOmega = \xi \be_x + \eta \be_y + \mu \be_z$.  If $\theta\in [0,\pi)$ and $\varphi\in [0,2\pi)$ are the polar and azimuthal angles, respectively, on the sphere (see \Cref{fig:spherical-coordinate}) and $\mu = \cos(\theta)$, then
\begin{align}\label{eq_polar_coordinates}
\begin{pmatrix}
    \xi \\ \eta \\ \mu
\end{pmatrix}
= 
\begin{pmatrix}
    \sin \theta  \cos \varphi  \\ \sin \theta  \sin \varphi \\ \cos \theta
\end{pmatrix}
= 
\begin{pmatrix}
    \sqrt{1-\mu^2}  \cos \varphi \\ \sqrt{1-\mu^2}  \sin \varphi \\ \mu
\end{pmatrix},
\end{align}
where $(\mu,\varphi) \in P = [0,1] \times [0,2 \pi)$.

The domain $X$ is assumed to be an extrusion of a set $D \subset \bbR^2$ in the $\be_z$ direction; that is, $X = D \times \bbR$ where $\partial D$ is Lipschitz (see \Cref{fig:D}).
Thus for any $\bx \in \p X$, $\bn(\bx) = n_x(x,y) \be_x + n_y(x,y) \be_y$. 
Furthermore, the material cross-sections, source, initial data, and boundary data are all asumed to be independent of $z$.  As a result, $\Psi$ is also independent of $z$. Moreover if $\Psi(\bx,\bsOmega,t)$ is a solution of the RTE, then so is $\Psi(\bx,\bsOmega - 2\mu \be_z,t)$. Thus by uniqueness of the RTE solution \cite{dautray2012mathematical_v6}, $\Psi$ must be an even function of $\mu$.

Motivated by the symmetry assumptions above, let $\Psi(\bx,\bsOmega,t) = \psi(x,y,\mu,\varphi,t)$, and let
\begin{align}
    \gamma^{\pm} = \{ (x,y,\mu,\varphi) \in \p D \times P \colon \pm (n_x \xi + n_y \eta) > 0 \}
\end{align}
parameterize the projection  of $\Gamma^\pm$ onto $\partial D \times \bbS^2_+$.  When expressed in terms of $\psi$, the RTE \eqref{eq_transport} becomes
\begin{subequations}
\label{eq_transport_2d}
	\begin{numcases}{}
  	\p_t \psi+ \xi \p_x \psi  + \eta \p_y \psi +\sigmat \psi 
        =  \frac{\sigmas}{2 \pi}\vint{\psi}_+ +Q, 
        & $(x,y,\mu,\varphi) \in D \times P, \; t >0$,
        \label{eq_transport_2d_eqn}\\
    \psi(x,y,\mu,\varphi,t) =  \psi^-(x,y,\mu,\varphi,t) 
    &   $(x,y,\mu,\varphi) \in \gamma^- , \; t>0$, \\
    \psi(x,y,\mu,\varphi,0) =  \psi_0(x,y,\mu,\varphi) 
    & $(x,y,\mu,\varphi) \in D \times P$,
 \end{numcases}
 \end{subequations}
where 
\begin{align}
    \vint{\psi}_+ = \int_{\bbS_+^2} \Psi(\bsOmega) d \bsOmega
    = \int_P  \psi(\mu,\varphi) d \mu d \varphi
    = \int^{1}_{0} \int_{0}^{2\pi} \psi(\mu,\varphi) d \mu d \varphi.
\end{align}
In particular all angular integrals in \eqref{eq_transport_2d} can be evaluated over $\bbS^2_+$.

\section{Benchmark problems}\label{sec_problems}

In this section, we formulate the two benchmark problems.  For each one, we provide (i) a detailed mathematical description of the domain and the problem inputs, (ii) a list of the quantities of interest to be computed, and (iii) a suggested list of design parameters to be used in later studies.

\subsection{Lattice benchmark}\label{sec_lattice}

The Lattice benchmark \cite{brunner2002forms, BRUNNER2005386} is characterized by a source that emits particles into a surrounding heterogeneous material environment.  The layout, material parameters, and a plot of the particle concentration from a high-resolution simulation are provided in \Cref{fig_lattice}.
The two-dimensional domain $D$ is a collection of three types of materials arranged in checkerboard pattern.  
The domain is initially devoid of particles, but over time particles enter the domain via an isotropic source that is positioned in the center of the domain.

\subsubsection{Mathematical description} 
\label{subsec_lattice_description}
The domain $D$ is the union of square blocks
\begin{equation}
    D = [-3.5,3.5] \times [-3.5,3.5] = \bigcup_{i,j=1}^7 \overline{C_{i,j}}.
\end{equation}
The blocks
\begin{equation}
 C_{i,j} 
 = (x_{i-1/2},x_{i+1/2}) \times (y_{j-1/2},y_{j+1/2})
\end{equation}
have edges coordinates
\begin{align}
    x_{i \pm 1/2} = x_i \pm \frac12 \dx
    \quad \text{and} \quad
    y_{j \pm 1/2} = y_j \pm \frac12 \dy,
\end{align}
center coordinates
\begin{align}
    x_i &= -3.0 + (i-1)\dx, \quad i = {1,\dots ,7},\\
    y_j &= -3.0 + (j-1)\dy, \quad j = {1,\dots ,7},
\end{align}
and thickness $\dx = \dy = 1$.

\begin{figure}

\begin{subfigure}[t!]{0.32\linewidth}
    \centering
    \includegraphics[width=0.9\textwidth,trim={0 0 6cm 0},clip]{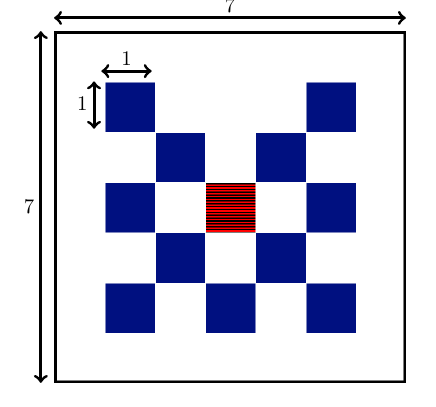}
    \vfill
    \caption{Geometric layout}
      \label{fig_lattice_layout}
\end{subfigure}
    \begin{subfigure}[t!]{0.32\linewidth}
     \vfill
    \centering
\begin{tabular}{l | c ccc}
subdomain &  $\sigma_t$ &  $\sigmas$ &$\sigmaa$ & $Q$ \\
\midrule 
white & $1$ & $1$ & $0$ & $0$\\
blue & $10$ & $0$ & $10$ & $0$ \\
red-striped & $1$ & $1$ &$0$ & $1$  \\
\bottomrule
        \end{tabular}   
    \caption{Material parameters}
    \label{fig_lattice_table}
\end{subfigure}
\begin{subfigure}[t!]{0.32\linewidth}
    \centering
 \includegraphics[width=\textwidth]{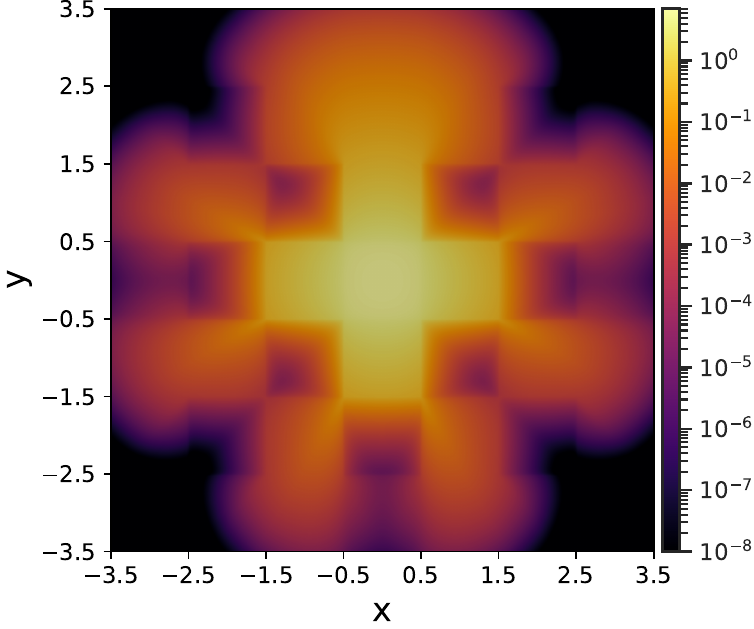}
    \vfill
    \caption{Reference solution}
    \label{fig_hohlraum_ref_sol}
\end{subfigure}
    \caption{The Hohlraum benchmark: (a) geometric layout; (b) material parameters; (c) approximation of the particle concentration $\phi = \vint{\psi}_+$ for a high-resolution simulation.  The simulation in (c) uses an order $N=22$  tessellation quadrature from \cite{JarrellAdams2011} (see also \Cref{subsubsec_angular} below) and a characteristic mesh length of $h=1\times 10^{-3}$, resulting in $1936$ angles and $16,904,270$ mesh cells in space. To reach the final time $T=3.2$ with a CFL number $0.7$ requires $9830$ time steps.}
    \label{fig_lattice}
\end{figure}

The blocks $C_{i,j}$ are collected into three subdomains with common material properties (c.f. \Cref{fig_lattice})
\begin{subequations}
\begin{alignat}{3}
    \text{(blue blocks):} &\,\,
    B &&= C_{2,2} \cup C_{4,2} \cup C_{6,2} 
    \cup C_{3,3} \cup C_{5,3}
    \cup C_{2,4} \cup C_{6,4} 
    \cup C_{3,5} \cup C_{5,5}
    \cup C_{2,6} \cup C_{6,6}  \\
    \text{(red blocks):} &\,\,
    R &&= C_{4,4} \\
    \text{(white blocks):} &\,\,
    W&&= \operatorname{int}(X \sim ({B \cup R})).
\end{alignat}
\end{subequations}
The cross-sections in each subdomain are given by  
\begin{align}
    \sigmaa = \sigmaa^B \chi_B
    \qquand
    \sigmas = \sigmas^W \chi_W + \sigmas^R \chi_R,
\end{align}
where $\sigmaa^B =10$, $\sigmas^W = \sigmas^R = 1$, and $\chi_{S}$ is the indicator function on any set $S \subset D$. The source is isotropic and uniform in magnitude over $R$, i.e.,
\begin{equation}
    Q = Q^R \chi_{R},
\end{equation}
where $Q^R = 1$.

The boundary and initial data needed for \eqref{eq_sn_2d} is $\psi^0 = 0$ and inflow $\psi^- = 0$. We simulate the solution until final time $T = 3.2$.
\subsubsection{Quantities of interest}
\label{subsubsec_lattice_QoI}

Let  $S_\ell =[-\ell,\ell] \times [-\ell,\ell]$ and let
\begin{align}
    \gamma_\ell^{\pm} = \{(x,y,\mu,
    \phi) \in \partial S_\ell\times P : \pm(n_x \xi + n_y \eta)  > 0\}.
\end{align}
be the outflow boundary for $S_\ell$.
Further, let 
       \begin{align}    
    \phi(x,y,t) = \int_P \psi(x,y,\mu,\phi,t) d\mu d\phi
    \end{align}
be the particle concentration.

We compute the following quantities of interest for the Lattice benchmark.
    \begin{itemize}
        \item\textbf{(QoI-1)}:
   the \textbf{outflow through the boundary at final time} $T$,  i.e., 
   \begin{align}\int_{\gamma_\ell^+} (n_x \xi + n_y \eta) \psi(x,y,\mu,\phi,T) ds(x,y) d \mu d \phi ,
   \end{align}
   for each $\ell \in \{1.5,2.5\}$.
 \item\textbf{(QoI-2)}:  the \textbf{total outflow over time}, i.e.,
    \begin{align}
        \int_0^T \int_{\gamma_\ell^+} (n_x \xi + n_y \eta) \psi(x,y,\mu,\phi,t) ds(x,y) d \mu d \phi dt ,
    \end{align}
  \item\textbf{(QoI-3)}: the \textbf{absorption at final time} $T$, i.e.,
    \begin{align}
         \int_{B} \sigmaa(x,y) \phi(x,y,T)  dxdy ,
    \end{align}

  \item\textbf{(QoI-4)}:   the \textbf{total absorption over time}, i.e.,
    \begin{align}\int_0^T \int_{B} \sigmaa(x,y) \phi(x,y,t)  dxdy dt , 
    \end{align}
      \item\textbf{(QoI-5)}:  the \textbf{particle mass at final time} $T$,  i.e.,
    \begin{align} \int_{D}  \phi(x,y,T)  dxdy .
    \end{align}
\end{itemize}

\subsubsection{Design Parameters}
In the current work, we compute solutions and quantities of interest using the default parameters give in \Cref{subsec_lattice_description}; see also \Cref{fig_lattice_table}. However, for future uncertainty quantification and design studies and for surrogate model training, we propose the use of material property values as design parameters. Specifically, we propose the following:
\begin{itemize}
 \item\textbf{(DP-1)}: the absorption cross-section $\sigmaa^B$ and the scattering cross-section $\sigmas^W$
 \item\textbf{(DP-2)}: the absorption cross-sections $\sigmaa^{{ij}}$ and scattering cross-sections $\sigmas^{{ij}}$ in each block $C_{i,j} \subset X $.
\end{itemize}
The parameter space for $\textbf{(DP-1)}$ is two-dimensional. Exploring this space will be challenging, but manageable, especially with the use of surrogates.  However, $\textbf{(DP-2)}$ introduces a parameter space with as many as $49$ dimensions. Strategies for exploring this space will need to be much more sophisticated.

\subsection{Hohlraum benchmark}\label{sec_hohlraum}

The Hohlraum benchmark is a cartoon of a real hohlraum, which is used in inertial confinement fusion devices.  The benchmark was originally proposed and simulated using a non-linear thermal radiative transfer model in \cite{brunner2002forms}.  A linearized version of the problem with fixed scattering and absorption cross-sections was introduced in \cite{hauck2013collision,osti_1550352,CROCKATT2019455,CROCKATT2020109765}. For the current benchmark we consider a symmetrized version of the linearized problem. 

\begin{figure}[h!]
\begin{subfigure}[t!]{0.32\linewidth}
    \centering    \includegraphics[width=1\textwidth,trim={3cm 11cm 22cm 0cm},clip]{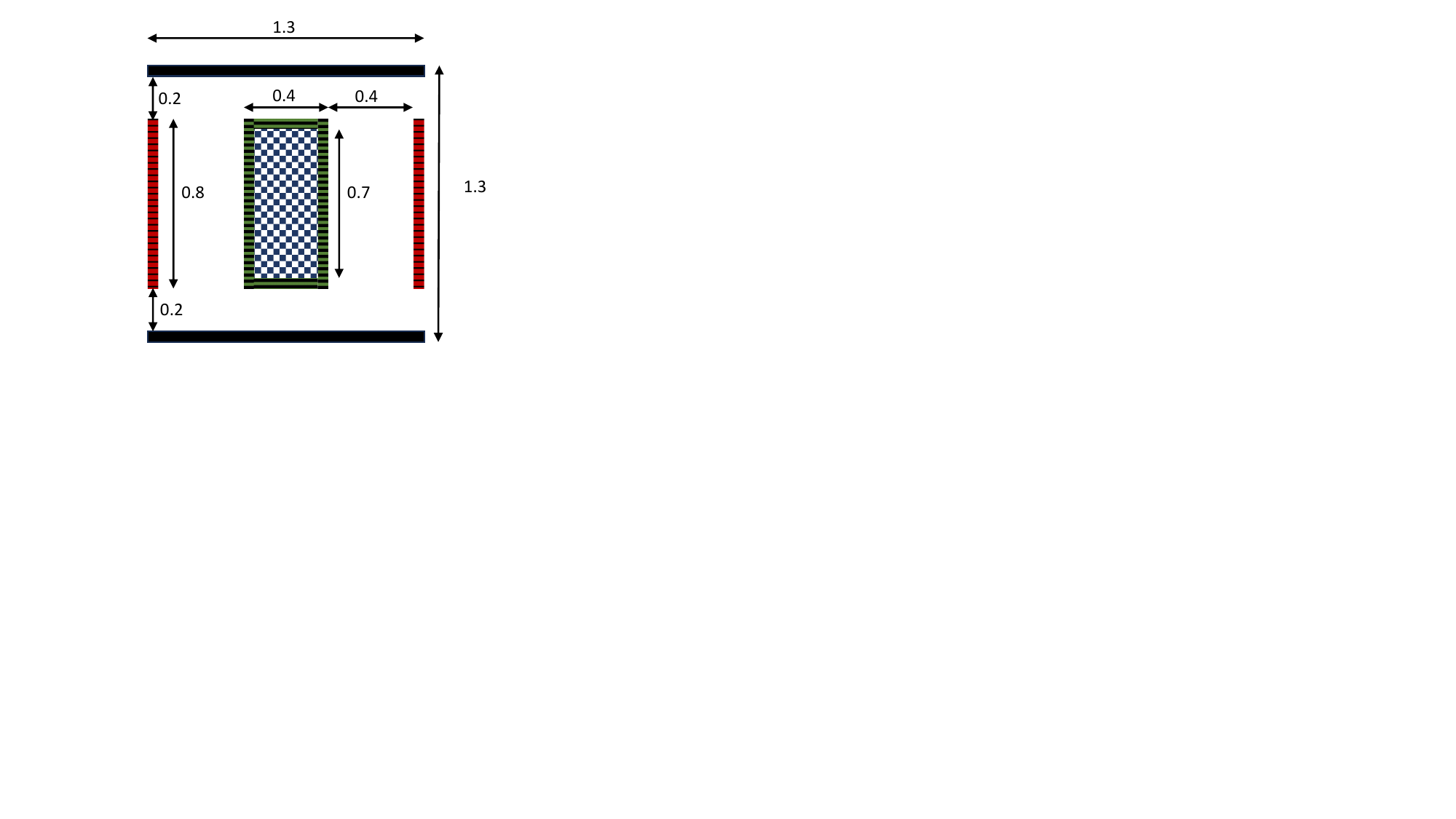}
    \vfill
    \caption{Geometric layout}
    \label{fig_hohlraum_schematic}
\end{subfigure}
        \begin{subfigure}[t!]{0.32\linewidth}
     \vfill
    \resizebox{!}{0.3\linewidth}{
\begin{tabular}{l | c ccc}
subdomain &  $\sigmat$ &  $\sigmas$ &$\sigmaa$ \\
\midrule 
white & $0.1$ & $0.1$ & $0$ \\
blue-checkered & $100$ & $0$ & $100$ \\
red & $100$ & $95$ &$5$ \\
green & $100$ & $90$ & $10$ \\
black & $100$ &$50$ & $50$ \\
\bottomrule
        \end{tabular}
         }
    \caption{Material parameters}
    \label{fig_hohlraum_table}
\end{subfigure}
\begin{subfigure}[t!]{0.32\linewidth}
\hspace{30pt}
 \includegraphics[width=\textwidth]{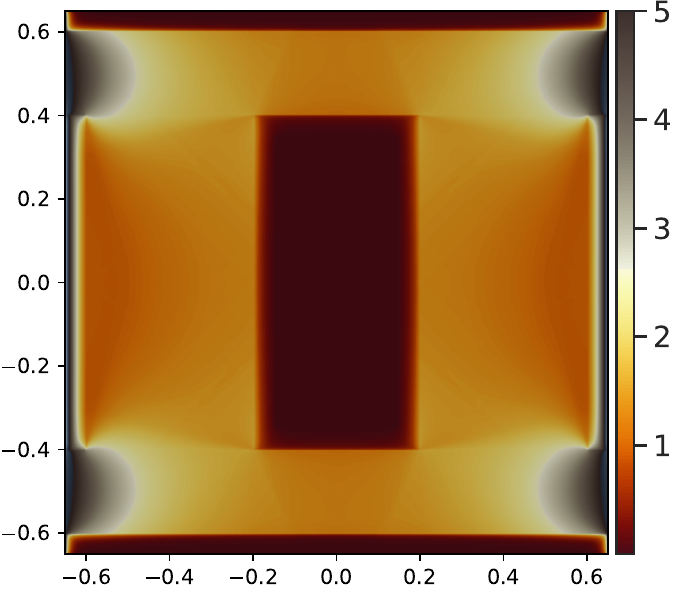}
    \vfill
    \caption{Reference solution}
    \label{fig_hohlraum_referece}
\end{subfigure}
    \caption{The Hohlraum benchmark: (a) geometric layout; (b) material parameters; (c) particle concentration $\phi = \vint{\psi}_+$ for a high-resolution simulation.  The simulation in (c) uses an order $N=22$  tessellation quadrature from \cite{JarrellAdams2011} (see also \Cref{subsubsec_angular} below) and a characteristic mesh length of $h=7.5\times 10^{-4}$, resulting in $1936$ angles and $7,121,004$ mesh cells in space. To reach the final time $T=2.6$ with a CFL number $0.8$ requires $18,405$ time steps.}
    \label{fig_hohlraum}
\end{figure}

The Hohlraum geometry consists of a square domain with shielding material on the top and bottom. In the center of the domain is a rectangular capsule with an outer later of material that is irradiated from the left and right sides of the domain, where additional shielding material limits the amount of radiation that directly hits the capsule.  The layout, material parameters, and a approximaiton of the particle concentration from a high-resolution simulation are given in \Cref{fig_hohlraum}.%
\footnote{The material properties in \Cref{fig_hohlraum_table} are the same as those reported in \cite{CROCKATT2019455,osti_1550352}.  However in \cite{CROCKATT2020109765} the parameter values of the black and blue regions were inadventently swapped in the text, but not in the actual simulations. Thus in subsequent work, \cite{schotthoefer2024structurepreservingneuralnetworksregularized, KROTZ2024113253}, the swapped values were used for simulations as well. In an attempt to unify the benchmark description, we urge the community to use the values in \Cref{fig_hohlraum_table}.}

\subsubsection{Mathematical description}
\label{subsec_hohlraum_description}
The spatial domain (see \Cref{fig_hohlraum_schematic}) is given by
\begin{align}
    \hspace{-2.0cm} D = [-0.65,0.65] \times [-0.65,0.65] = \overline K \cup \overline R \cup \overline G \cup \overline B \cup \overline W.
\end{align} 
The  set of black blocks is denoted by
$K= K_1 \cup K_2 $, where
    \begin{align}
        K_1 = (-0.65,0.65) \times (-0.65,-0.60),
    \quad \text{and} \quad 
    K_2 = (-0.65,0.65)  \times (0.60,0.65).
    \end{align}
The set of red blocks is denoted by
$R  = R_1 \cup R_2$, where
        \begin{align}
    R_1 = (-0.65,-0.60) \times (-0.40,0.40),
    \quad \text{and} \quad 
    R_2 = (0.60,0.65) \times (-0.40,0.40).
      \end{align}
The set of green blocks is denoted by
$ G= G_1 \cup G_2 \cup G_3 \cup G_4$, where the blocks on the left and right of the capsule are 
    \begin{align}
    G_1 = (-0.20,-0.15) \times (-0.35,0.35),
    \quad \text{and} \quad 
    G_2 =  (0.15,0.20) \times (-0.35,0.35),
   \end{align}
   and the blocks on the top and bottom of the capsule are
    \begin{align}
    G_3 = (-0.20,0.20) \times (-0.40,-0.35),
    \quad
    G_4 = (-0.20,0.20) \times (0.35,0.40).
    \end{align}
    Lastly, the set of white blocks is denoted by $W = \operatorname{int}(X \sim ({K \cup R \cup B \cup B }))$.

Again let $\chi_S$ be the indicator on a set $S$.  Then the absorption and scattering cross-sections, as illustrated in \Cref{fig_hohlraum_schematic}, are given by
\begin{align}
    \sigmaa &=  \sigmaa^W \chi_W +\sigmaa^B \chi_B +\sigmaa^R \chi_R + \sigmaa^G \chi_G +  \sigmaa^K \chi_K,   
        \end{align}
where
\begin{equation}
\sigmaa^W = 0,  
\quad 
\sigmaa^B = 100,
\quad
\sigmaa^R = 5, 
\quad
\sigmaa^G = 10,
\quad 
\sigmaa^K = 50. 
\end{equation}
 and
 \begin{align}
    \sigmas &= \sigmaa^B \chi_B +\sigmaa^R \chi_R + \sigmaa^G \chi_G +  \sigmaa^K \chi_K,
\end{align}
where
\begin{equation}
\sigmas^W = 0.1,  
\quad 
\sigmas^B = 0,
\quad
\sigmas^R = 95, 
\quad
\sigmas^G = 90,
\quad 
\sigmas^K = 50. 
\end{equation}

There are no initial particles and no source in the Hohlraum benchmark.  Thus $Q = 0$ and $\psi_0 = 0$ in \eqref{eq_sn_2d}.
Meanwhile, the inflow data in \eqref{eq_sn_2d} is given by
\begin{align}
\psi^-(x,y,\mu,\varphi) =
\begin{cases}
1.0, \quad & x=-0.65, \;|y| < 0.65,\; \xi > 0 ,\\
1.0, \quad & x=0.65,\quad |y|< 0.65,\; \xi < 0 ,\\
0, & \text{otherwise},
\end{cases}
\end{align}
where $\xi = \sqrt{1-\mu^2} \cos \varphi$ is defined in \eqref{eq_polar_coordinates}.  The final simulation time is $T = 2.6$.

\subsubsection{Quantities of interest}
\label{subsubsec_hohlraum_QOI}
Recall that
       \begin{align}    
    \phi(x,y,t) = \int_P \psi(x,y,\mu,\phi,t) d\mu d\phi
    \end{align}
is the particle concentration.  Further let 
\begin{equation}
    J_x = \int_P \xi \psi(x,y,\mu,\phi,t) d\mu d\phi
    \quand
    J_y = \int_P \eta \psi(x,y,\mu,\phi,t) d\mu d\phi
\end{equation} be the 
$x$ and $y$ currents respectively.

We are interested in measuring the following quantities of interest for the Hohlraum benchmark:

\begin{itemize}
    \item \textbf{(QoI-1)}: The \textbf{space-time average of the particle concentration and current near  the center of the channels } is given by 
\begin{align}
    &\frac{1}{C_{n,p,r}}\int_{T_{n-1}}^{T_{n} }  \int_{B_r(x_p,y_p)}\phi (x,y,t) dx dy,
    \\
    &\frac{1}{C_{n,p,r}}\int_{T_{n-1}}^{T_{n} }   \int_{B_r(x_p,y_p)} J_x(x,y,t) dx dy,
    \quand
   \frac{1}{C_{n,p,r}} \int_{T_{n-1}}^{T_{n} }  \int_{B_r(x_p,y_p)} J_y(x,y,t) dx dy 
\end{align}
where $T^n = n T/ 10$ for $n \in \{0, \dots, 10\}$, the probe ${B_r(x_p,y_p)}$ is a ball with center $(x_p,y_p)$ and radius $r=0.01$, and $C_{n,p,r}$ is the space-time volume of the set $B_r(x_p,y_p) \times (T_n - T_{n-1})$. There are four probes centered in the channels on each side of the capsule: \begin{equation}
    (x_1,y_1) = (-0.4,0),
    \quad
    (x_2,y_2) = (0.4,0),
    \quad
    (x_3,y_3) = (0,-0.5),
    \quad
    (x_4,y_4) = (0, 0.5).
\end{equation}
 \item \textbf{(QoI-2)}:
the \textbf{total particle absorption the various materials} is given by 
   \begin{align}\int_0^T \int_{S}  \sigmaa(x,y) \phi(x,y,t)  dxdy dt 
   \end{align}
where $S \in {G \cup B, R, K}$.
 \item \textbf{(QoI-3)}: the \textbf{average particle absorption given a blockwise partition of the green domain}, specified by the $N_G = 44$ green blocks of size $0.05 \times 0.05$, denoted by $g_{i}$. The blocks are enumerated counter-clockwise, starting at the botton left block with center $(x,y)=(-0.175,-0.375)$.  Then
\begin{align}
    A_G 
    = \frac{1}{N_G} \sum_{i} A_{i},
\quad\text{where}\quad
      A_{i}
    = \frac{1}{\operatorname{vol}(g_{i})}  \int_{g_{i}}  \int_0^T\sigmaa(x,y) \phi(x,y,t) dt dxdy.
\end{align}
 \item \textbf{(QoI-4)}:
the \textbf{variation in the particle absorption given the blockwise partition of the green domain}:
\begin{align}
V_G = \frac{1}{N_G}
\sum_{i}  |A_{G} - A_{i}|^2 .
\end{align}
 \item\textbf{(QoI-5)}:  the \textbf{average particle absorption given a line $L$ through the center of each green block  $g_{i,j}$}, 
 \begin{align}
     A_L = \frac{1}{\operatorname{length}(L)} 
     \int_{L} \int_0^T \sigmaa(x,y) \phi(x,y,t)(t) dt ds(x,y).
  \end{align}
 \item\textbf{(QoI-6)}:  the \textbf{variation in the particle absorption given a line $L$ through the center of each green block  $g_i$}, 
 \begin{align}
 V_L&= \frac{1}{\operatorname{length}(L)}
 \int_{L}    |A_L - \int_0^T\sigmaa(x,y) \phi(x,y,t)dt|^2  ds(x,y).
 \end{align}
\item\textbf{(QoI-7)}:    and the \textbf{particle mass at final time}, i.e.,
    \begin{align} \int_{D}  \phi(x,y,T)  dxdy .
    \end{align}
\end{itemize}

The pairs \textbf{QoI-3/4} and \textbf{QoI-5/6} both attempt to assess the uniformity of the absorption around the surface of the capsule, but in slightly different ways.

\subsubsection{Design parameters}
In the current work, we compute solutions and quantities of interest using the default parameters given in \Cref{subsec_hohlraum_description}; see also \Cref{fig_hohlraum_table}. However, for future uncertainty quantification and design studies and for surrogate model training, we propose the use of geometric design parameters. Specifically, we propose the following:

\begin{itemize}
   \item \textbf{(DP-1)}: the location of the top and bottom edges of the red strips on each side,
   \item \textbf{(DP-2)}: the thickness of the red strips, via location of the interior edge,
    \item\textbf{(DP-3)}: the variation of the capsule from the center.
\end{itemize}
The scenario \textbf{(DP-1)} involves as many as four parameters, \textbf{(DP-2)} as many as two parameters, and \textbf{(DP-3)} just one parameters.  It is also possible to variable the material cross-sections as in the Lattice benchmark.  However since the current nominal values are estimated from the associated nonlinear thermal radiative transfer model, we contend that it is more physically meaningful to vary the material parameters as part of a nonlinear simulation.

\section{Numerical discretization}\label{sec_numerics}

In this section, we present the discretization that is used to compute numerical solutions of \eqref{eq_transport_2d}.  The discretization methods are all well-known, but the details of the presentation are provided for completeness.

\subsection{Angular discretization: the \sn equations }
The discrete ordinates method \cite{carlson1968transport,azmy2010advances}, commonly referred to as the \sn method, is a well-known and widely used technique for angular discretization.  As the name suggests, the method approximates the transport solution on a finite set of ordinates, which are then used to approximate angular integrals.  While the ordinates can be chosen from any suitable quadrature rule on the sphere, the approximation properties of the \sn solution often depend heavily on that choice.

%
%
We apply the \sn discretization directly to the reduced problem \eqref{eq_transport_2d}.  Given a family of quadrature formulas indexed by an integer $N$, let $M = M(N)$ be the number of quadrature points on $\bbS_+^2$ of the form
\begin{equation}
    \bsOmega_k = \xi_k \be_x + \eta_k \be_y + \mu_k \be_z,
    \quad k=1, \dots, M .
\end{equation}
where $\mu_k \in [0,1]$ and the components $\xi_k$ and 
$\eta_k$ are defined in terms of $\mu_k$ and the angle $\varphi_k \in [0,2\pi)$ via \eqref{eq_polar_coordinates}. Let $\{w_k\}_{k=1}^{M}$ be corresponding set of weights, normalized such that $\sum_{k=1}^{M} w_k = 2 \pi$, and let
\begin{equation}
   \p D^{\pm}_k = \{ (x,y) \in \p D : \pm (n_x \xi_k + n_y \eta_k) > 0 \} \subset \p D
\end{equation}
be the inflow (-) and outflow (+) boundaries associated to each $\bsOmega_k$.
The \sn approximation of \eqref{eq_transport_2d} is:  \textit{Find $\bspsi^N(x,y,t) = [\psi_1(x,y,t), \dots ,\psi_{M}(x,y,t)]^\top$ such that}
\begin{subequations}
    \label{eq_sn_2d}
\begin{numcases}{\hspace{-20pt}}
   	\p_t \psi_k+ \xi_k \p_x \psi_k  + \eta_k \p_y \psi_k +\sigmat \psi_k 
        =  \frac{\sigmas}{2\pi}\sum_{l=1}^{M} w_l \psi_l + Q,  
         &$(x,y) \in D, \; 1 \leq k \leq M,  \; t >0$,  
         \label{eq_sn_2d_eqn}\\
     \psi_k(x,y,t) =  \psi^-(x,y,\mu_k, \varphi_k,t) 
     &$(x,y) \in \p D_k,  \;1\leq k \leq M, \; t > 0,$ \\
     \psi_k(x,y,0) =  \psi_0(x,y,\mu_k, \varphi_k)
     &$(x,y) \in D , \; 1\leq k \leq M$.
  \end{numcases}
  \end{subequations}
The \sn equations in \eqref{eq_sn_2d} form a system of $M$ linear, time-dependent, hyperbolic conservation laws in two space dimensions.


\subsection{Spatial discretization}
We partition the spatial domain $D$ of the \sn system \eqref{eq_sn_2d} using an unstructured mesh $\cT_h$ consisting of $N_h$ open triangular cells $C_i$, $i \in \{ 1, \dots, N_h\}$, with area $A_i$ that is  controlled by a parameter $h >0$ and centroids $\bm{x}_{i}$. The parameter $h$ denotes the maximum edge length of the mesh, which is constructed with a Delaunay triangulation algorithm.  We assume that the mesh is constructed so that the material parameters and sources are piece-wise constant on each mesh cell.

For each $i$, let
\begin{equation}
   e_{i,j} = \overline{C_i} \cap \overline{C_j} ,
   \quad 
   j \in \{ 1, \dots, N_h\}
\end{equation}
be the edges shared by $C_i$, each with length $\ell_{i,j}$, and let 
\begin{equation}
   \bar{S}_{i} = S_{i} \cup i 
   \quad\text{where}\quad S_{i} = \{ j: \ell_{i,j} > 0 \} 
\end{equation}
be the stencil for $C_i$.   
Associated to each edge $e_{i,j}$ is the midpoint $\bm{x}_{i,j}$ and a unit vector $\bm{n}_{i,j}$ normal to $e_{i,j}$ that points from $C_i$ into $C_j$.

To compactify the notation, we write \eqref{eq_sn_2d_eqn} in the form

\begin{equation}
\label{eq_sn_2d_eqn_compact}
    \p_t \psi_k + \bsomega \cdot \grad_{\bm{x}}
    \psi_k + \sigmat \psi_k =  \frac{\sigmas}
    {2 \pi} \sum_{l=1}^M w_l \psi_l + Q_k,    
\end{equation}
where $\bm{x} = x \be_x + y \be_y$ and $\bsomega = \xi \be_x + \eta \be_y$ are understood as vectors in $\bbR^2$.  A semi-discrete finite volume method for \eqref{eq_sn_2d_eqn_compact} evolves the approximate cell averages
\begin{equation}
    \psi_{k,i}(t) \approx \frac{1}{A_i}\int_{C_i} \psi_{k}(\bm{x},t) d \bm{x}.
\end{equation}
It takes the form
\begin{equation}
\label{eq_sn_2d_eqn_fv}
    A_i \p_t \psi_{k,i} + \sum_{j \in S_{i}}  (\bsomega \cdot \bm{n}_{i,j}) \ell_{i,j} \hat{\psi}_{k,i,j} + \sigmati \psi_{k,i} =  \frac{\sigmasi}
    {2 \pi} \sum_{l=1}^M w_l \psi_{l,i} + Q_{k,i},
\end{equation}
where $\sigmati$ and $\sigmasi$ are the (constant) values of $\sigmat$ and $\sigmas$ on $C_i$, respectively, and $\hat{\psi}_{k,i,j}$ is the upwind numerical trace for the edge $e_{i,j}$, i.e., 
\begin{align}\label{eq_upwind1stOrder}
    \hat{\psi}_{k,i,j}(t)
    = R_{k,i}(t) \cH(\bsomega_k \cdot \bm{n}_{i,j}) + R_{k,j}(t)(1-\cH(\bsomega_k \cdot \bm{n}_{i,j}).
\end{align}
Here $\cH$ is the Heaviside function and $R_{k,i}$ is an approximation of $\psi_k$ at the point $\bm{x}_{i,j}$: 
\begin{equation}
    R_{k,i}(t) = \psi_{k,i}(t) + L_{k,i} \bm{g}_{k,i}(t) \cdot \bm{r}_{i,j},
\end{equation}
where $\bm{r}_{i,j} = \bm{x}_{i,j} - \bm{x}_i$, $\bm{g}_{k,i}$ approximates the gradient of $\psi_k$ in $C_i$:
\begin{equation}
   \bm{g}_{k,i}(t) = \frac{1}{A_i} \sum_{j \in S_{i}} \ell_{i,j} \bm{n}_{i,j} 
   \left(\frac{\psi_{k,i}(t) + \psi_{k,j}(t)}{2}\right),
\end{equation}
and $L_{k,i}$ the Venkatakrishnan slope limiter. 

The Venkatakrishnan limiter \cite{Blazek2015,venkat_limiter} is given by 
\begin{align}
\label{eq_V_limiter}
\begin{aligned}
    L_{k,i} = \min_{j\in S_{i}}
        \begin{cases}
    \frac{1}{\Delta}\left(\frac{\left(\beta_{k,i,\textup{max}}^2 +\epsilon^2\right)\Delta + 2\Delta^2\beta_{k,i,\textup{max}}}{\beta_{k,i,\textup{max}}^2 + 2\Delta^2 + \beta_{k,i,\textup{max}}\Delta + \epsilon^2} \right) \qquad \textup{ if } \Delta,>0\\
  \frac{1}{\Delta}\left(\frac{\left(\beta_{k,i,\textup{min}}^2 +\epsilon^2\right)\Delta + 2\Delta^2\beta_{k,i,\textup{min}}}{\beta_{k,i,\textup{min}}^2 + 2\Delta^2 + \beta_{k,i,\textup{min}}\Delta + \epsilon^2} \right) \qquad\,\, \textup{ if } \Delta<0,\\
  1 \qquad\qquad\qquad\qquad\qquad\qquad\qquad\,\,\,\,\, \text{ if } \Delta=0,
\end{cases}
\end{aligned}
\end{align}
where \begin{align}
\begin{aligned}
\Delta = \frac{1}{2} \bm{g}_{k,i}\cdot \bm{r}_{i,j}.
    \end{aligned}
\end{align}
The addition auxiliary variables in \eqref{eq_V_limiter} are given by
\begin{align}
    \beta_{k,i,\textup{max}} =\psi_{k,i,\textup{max}}(t)-\psi_{k,i} \qquad\textup{and}\qquad
    \beta_{k,i,\textup{min}} =\psi_{k,i,\textup{min}}(t)-\psi_{k,i},
\end{align}
where
\begin{align}
\begin{aligned}
    \psi_{k,i,\textup{max}}=\max_{j\in \bar{S}_{i}}
    \left\{\psi_{k,j}(t)\right\}
    \quand 
    \psi_{k,i,\textup{min}}=\min_{j\in \bar{S}_{i}}
    \left\{\psi_{k,j}\right\},
    \end{aligned}
\end{align}
and $\epsilon = KA_i^3$, typically with $K\in\mathcal{O}(1)$. In this work, we set $K=1$.  

The Venkatakrishnan limiter is a smooth version of the Barth-Jespersen limiter \cite{Blazek2015}, the latter of which is positivity preserving and total variation diminishing. The Venkatrakrishnan limiter is only total variation bounded, but less diffusive. We have found that the Venkatrakrishnan limiter performs well in the test cases considered in this paper. 

Boundary conditions are implemented using ghost cells. For each cell $C \subset D$ with an edge $e$ on the domain boundary, we associate a ghost cell with the same area that is generated by reflecting $C$ over the boundary.  This ghost cell is assigned a constant numerical value that the equal to the inflow value $\psi^-$ at the center of the edge $e$.   Fluxes at the domain boundary are computed using first-order upwind flux. 

\subsection{Temporal discretization}

Let $\bspsi$ be an time dependent matrix with components $[\bspsi(t)]_{i,k} = \psi_{i,k}(t)$.  In terms of $\bspsi$, the equations in \eqref{eq_sn_2d_eqn_fv} can be written in the abstract form 
\begin{equation}
\label{eq_ODE}
    \dot{\bspsi} = \bF(\bspsi).
\end{equation} 
which can be evolved in time using a SSP-RK2 method (i.e., Heun's method).  Given a final solution time $T$, we divide $[0,T]$ into a uniform mesh with $N_t$ cells of size $\dt = T/N_t$.  Given the vector $\bspsi^n \approx \bspsi(n\dt)$, $n \in \{0,\dots,N_t-1\}$, Heun's method computes $\bspsi^n \approx \bspsi((n+1)\dt)$ as follows:
\begin{align}
   \bspsi^{(1)} = \bspsi^{n} + \dt \bF(\bspsi^{n}), \qquad
   \bspsi^{(2)} = \bspsi^{(1)} + \dt \bF(\bspsi^{(1)}), \qquad
   \bspsi^{n+1} = \frac{\bspsi^{n} + \bspsi^{(2)}}{2},
\end{align}
where $\bspsi^{0} = \bspsi(0)$.

\section{Implementation}\label{sec_implementation}
The discrete ordinate, finite volume numerical scheme is implemented in KiT-RT \cite{kusch2023kit}\footnote{Software and source code are available at \url{https:/github.com/KiT-RT/kitrt_code}. The results are obtained using commit \url{59e24a10d73e57a7872aa128539c1d31a5087485}}, an open source \texttt{C++} solver platform for radiation transport.  All simulations presented use KiT-RT with the Python wrapper CharmKiT\footnote{Software and source code are available at \url{https:/github.com/ScSteffen/CharmKiT}. The results are obtained using commit \url{56935deba165d83b00775934e840ab49b7528405}}. All simulation logs, solver settings, solution flow-fields, and time-series are available at DOI \url{10.13139/OLCF/2540565}.

\subsection{Solver settings}

\subsubsection{Angular quadrature} 
\label{subsubsec_angular}

\begin{figure}
    \centering
    \includegraphics[width=0.5\linewidth]{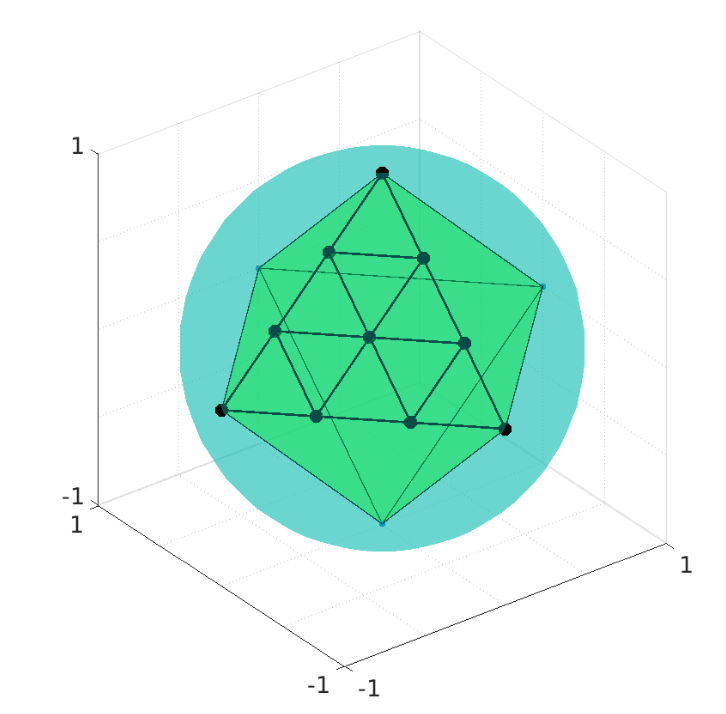}
    \caption{Octahedron for the tessellation quadrature. The image is obtained from \cite{Camminady2021}[Figure 2.7a] Each face of the octahedron is triangulated. The centroids of the triangles are then  computed and mapped to the unit sphere to form the quadrature points. The area of each projected triangle is calculated using the spherical excess and constitutes the quadrature weight.}
    \label{fig_octa}
\end{figure}

We choose a tessellation quadrature \cite{JarrellAdams2011}. 
This quadrature is created by constructing an octahedron with vertices on the unit sphere. Each face of the octrahedron is then triangulated. 
The centroids of the triangles are then  computed and mapped to the unit sphere to form the quadrature points. The area of each projected triangle is calculated using the spherical excess and constitutes the quadrature weight. The octahedron and the triangulation of one face is illustrated in \Cref{fig_octa}. For this choice, the number of quadrature points for $\mathbb{S}_+^2$ is given by $M(N)=4N^2$. The tessellation quadrature exhibits fewer ray effects than other quadratures, such as tensorized Gauss-Legendre \cite{walters1987use,atkinson1982numerical} with similar angular resolution.  Ray effects are a well-known numerical artifact in \sn discretizations that dissipate as the angular resolution is increased.  We discuss them in more detail below.

\subsubsection{Design of the spatial grids}\label{sec_lattice_grid}

We use unstructured, triangular grids that are designed to fit the material geometry in each benchmark. The grids are generated using \texttt{gmsh} \cite{geuzaine2009gmsh}.

\begin{figure}
\begin{subfigure}[t]{0.52\textwidth}
        \centering
    \includegraphics[ width=0.95\textwidth]{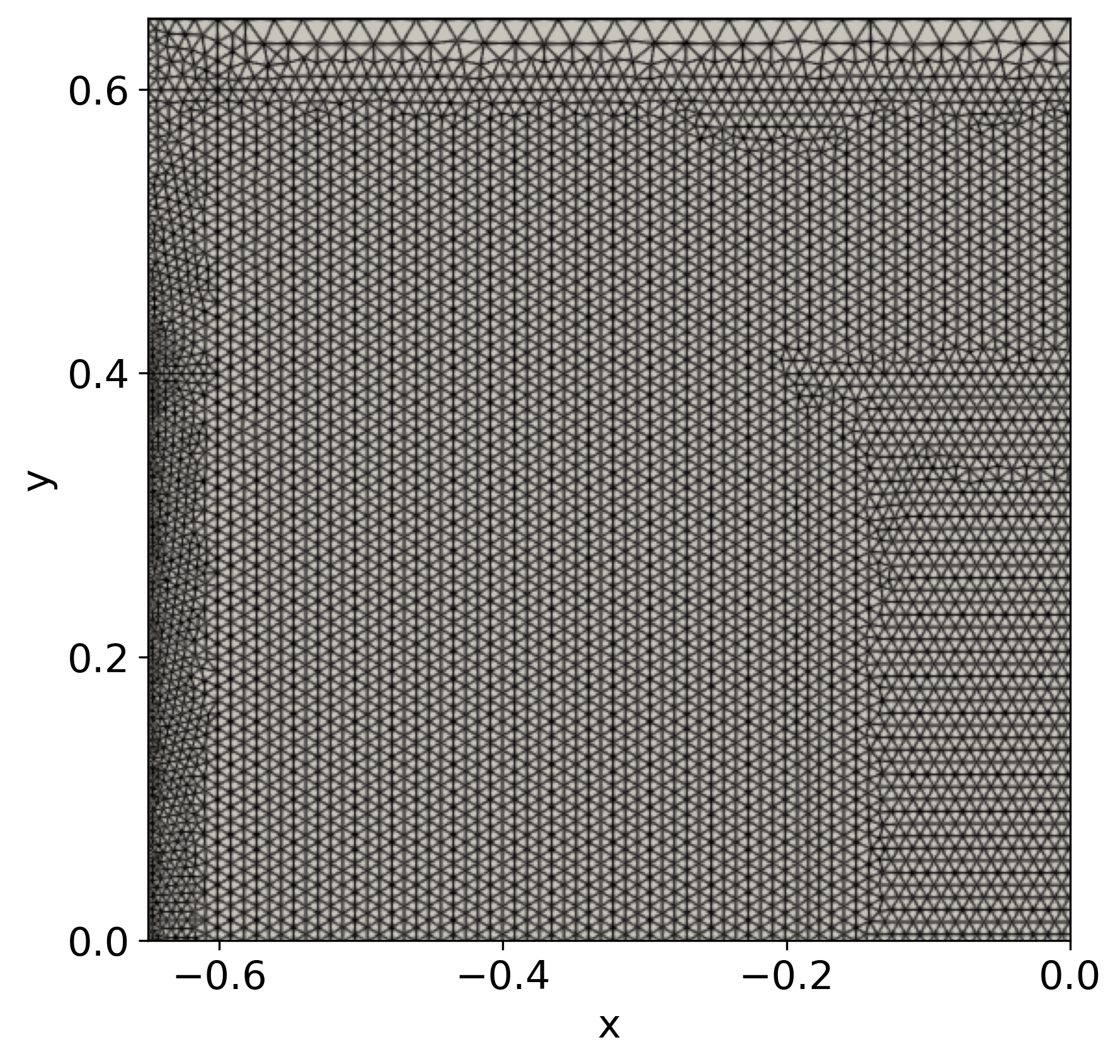}
    \vfill
    \caption{Hohlraum benchmark:  Top left quarter of the domain with mesh parameter $h=0.05$.  The mesh is symmetric with respect to the $x$ and $y$ axes.}
    \label{fig_mesh_hohlraum_unstructured}
\end{subfigure}
\hspace{30pt}
\begin{subfigure}[t]{0.43\textwidth}
        \centering
    \includegraphics[ width=0.7\textwidth]{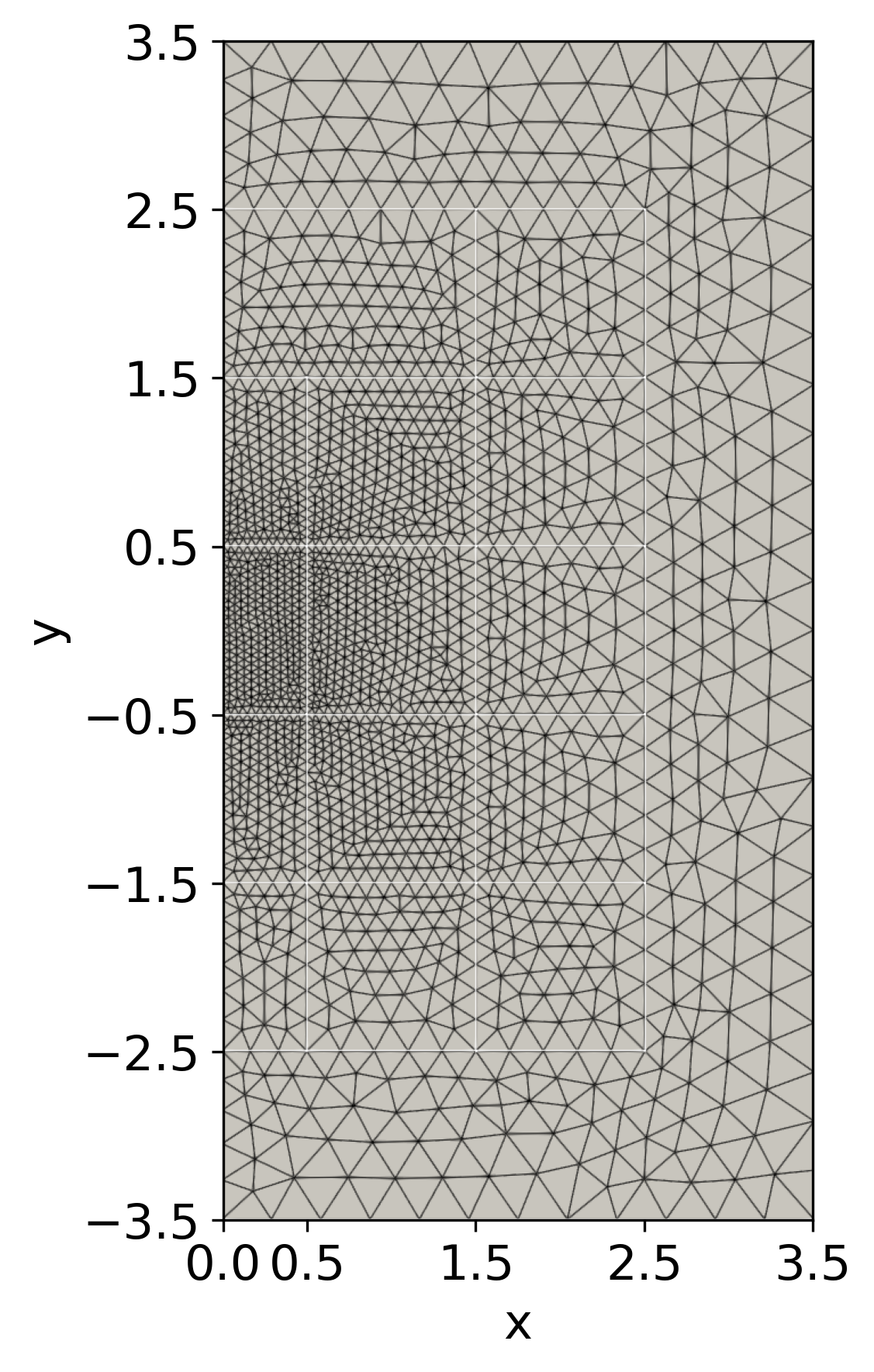}
    \vfill
    \caption{Lattice benchmark:  Right half of the domain with mesh parameter $h=0.05$.  The mesh is symmetric with respect to the $y$ axis.}
    \label{fig_mesh_half_lattice_unstructured}
\end{subfigure}
\caption{Unstructured triangular meshes for the two benchmarks.  The meshes are generated with the \texttt{gmsh} software \cite{geuzaine2009gmsh}}
\end{figure}

For the Lattice benchmark, we mesh $B$, $W$, and $R$ independently so that cells edges are perfectly aligned to material interfaces. In addition, the mesh is designed to respect the symmetry of the domain with respect to the $y$ axis. Because the solution is expected to decrease sharply at the interface wiyh the inner four blue squares, we increase the mesh resolution in this area while using coarser cells further away from the central source.  More precisely, the characteristic length $h_i$, which defines the upper limit for the edge length of the mesh cells in a region, is given by
\begin{align}
h_i = 
\begin{cases}
 h, & \quad \norm{\bm{x}}_\infty \leq 1,\\
2  h,& \quad 1< \norm{\bm{x}}_\infty \leq 2,\\
 4  h,& \quad 2< \norm{\bm{x}}_\infty \leq 2.5,\\
 6  h,& \quad 2.5<\norm{\bm{x}}_\infty \leq 3.5.
\end{cases}
\end{align}
The right half of the Lattice mesh for $h=0.05$ is displayed in \Cref{fig_mesh_half_lattice_unstructured}.

For the Hohlraum benchmark, the grid is designed similarly so that the mesh cells are aligned with the boundaries of the subdomains. Although the problem is symmetric with respect to $x$ andn $y$ axes, we do not enforce the mesh to be so because it complicates the probe measurements associated with \textbf{QoI-1}.
As the numerical solution is expected to decay sharply on the inner side of the shielding material on the left- and right-hand side of the domain, we refine the grid there. Meanwhile, because the resolution in the top and bottom shielding is expected to be less important for overall simulation quality, we coarsen in these regions.  More precisely, the characteristic length $h_i$ is given by
\begin{equation}
h_i = 
\begin{cases}
 0.5h, & \quad \bm{x} \in R, \\
 h, & \quad \bm{x} \in D \sim  (R \cup K), \\
 2h, & \quad \bm{x} \in K, \\
\end{cases}
\end{equation}
where $D$ is the domain and $R$ and $K$ are defined in \Cref{sec_hohlraum}.
The upper left quadrant of the mesh is given in \Cref{fig_mesh_hohlraum_unstructured}. 

\subsubsection{Time step}

The time step is set to $\dt = \text{CFL} \cdot \dx$, where $\dx = \min_{1\leq i \leq N_h} \sqrt{A_i}$. For the Lattice benchmark, $\text{CFL} = 0.7$, and for the Hohlraum benchmark $\text{CFL} = 0.7$.

\subsection{Parallelization strategy}
We employ a hybrid parallelization scheme using distributed memory parallelism with MPI and shared memory parallelism with OpenMP. The assumption of isotropic scattering limits coupling between discrete angles of the \sn system to the quadrature sum $\sum_{l=1}^{M} w_l \psi_l$ in \eqref{eq_sn_2d_eqn}.  Thus we parallelize the angular discretization with distributed memory using MPI. Given a quadrature with $M$ discrete angles and an allocation with $N_{\rm{MPI}}+1$ MPI nodes, node $p$ holds a partition of the \sn system containing 
$M_p= \left\lfloor \frac{M}{N_{\rm{MPI}}} \right\rfloor
$ angles. Processor $N_{\rm{MPI}}+1$ receives the remainding $M \textup{ mod } N_{\rm{MPI}} $ angles.

The computation of the quadrature sum in \eqref{eq_sn_2d_eqn} requires two all-reduce operations per time-step (once for each stage of Heun's method).  We parallelize operations on each node using shared memory parallelization via OpenMP.  The resulting shared memory communication pattern is typical for finite volume schemes, where neighboring mesh cell s of different OpenMP threads require communication for evaluating numerical fluxes.
Lastly, the partitioned angular flux on the decomposed domain of a single OpenMp thread is parallelized with SIMD parallelism for fine-grained arithmetic operations. 

All simulations for this publications are run on the CADES Baseline cluster of the National Center for Computational Sciences at the Oak Ridge National Laboratory. We use $N_{\rm{MPI}}+1=41$ MPI nodes equipped with 2X AMD 7713 processors with $128$ OpenMP threads.

\section{Simulation results}\label{sec_results}

In this section, we present the simulation study for the Lattice \cite{brunner2002forms} and Hohlraum \cite{CROCKATT2020109765} benchmarks, using the \sn discretization and  second-order finite-volume scheme described in the previous section.  

\begin{figure}[h!]

\begin{subfigure}[t!]{0.32\linewidth}
  \centering
    \includegraphics[height=5.cm]{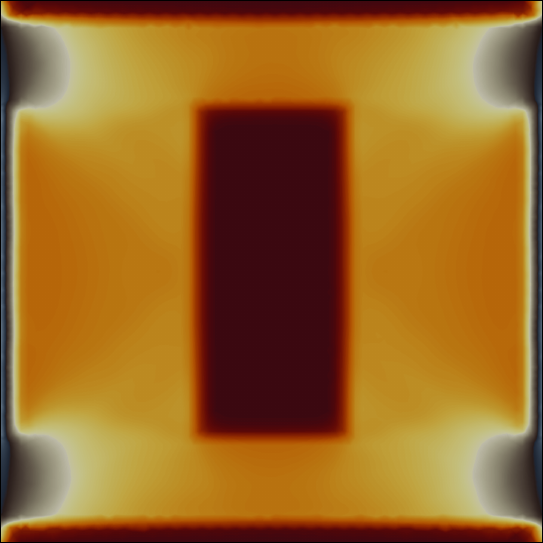}
    \vfill
    \caption{$h=0.02$}
\end{subfigure}
\hfill
\begin{subfigure}[t!]{0.32\linewidth}
  \centering
    \includegraphics[height=5.cm]{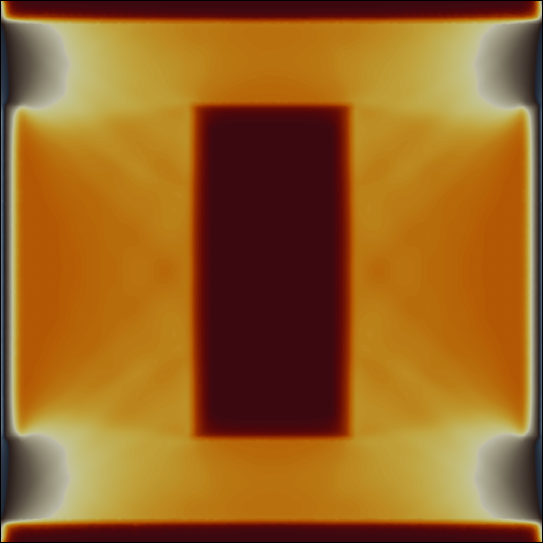}
    \vfill
    \caption{$h=0.01$}
\end{subfigure}
\hfill
\begin{subfigure}[t!]{0.32\linewidth}
  \centering
    \includegraphics[height=5.05cm]{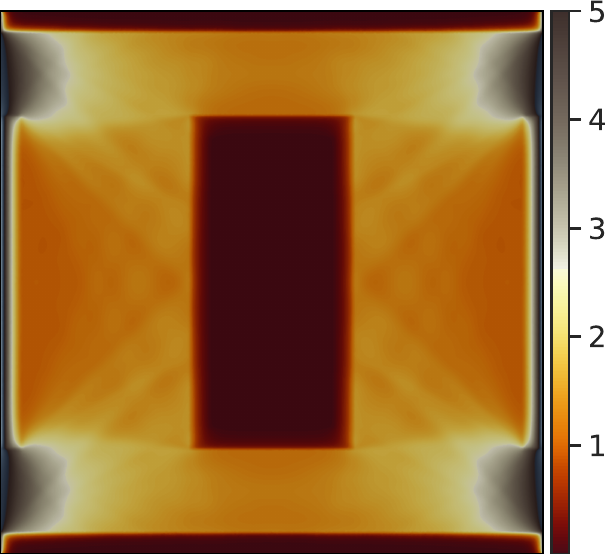}
    \vfill
    \caption{$h=0.005$}
\end{subfigure}
\begin{subfigure}[t!]{0.32\linewidth}
  \centering
    \includegraphics[height=5.cm]{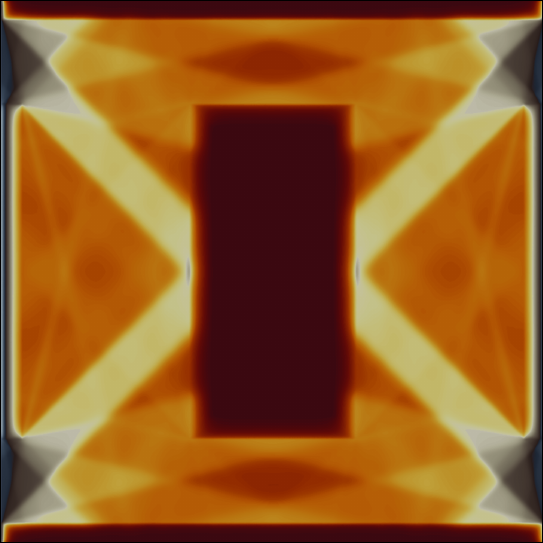}
    \vfill
    \caption{S$_2$}
    \label{fig_hohlraum_S_4}
\end{subfigure}
\hfill
\begin{subfigure}[t!]{0.32\linewidth}
  \centering
    \includegraphics[height=5.cm]{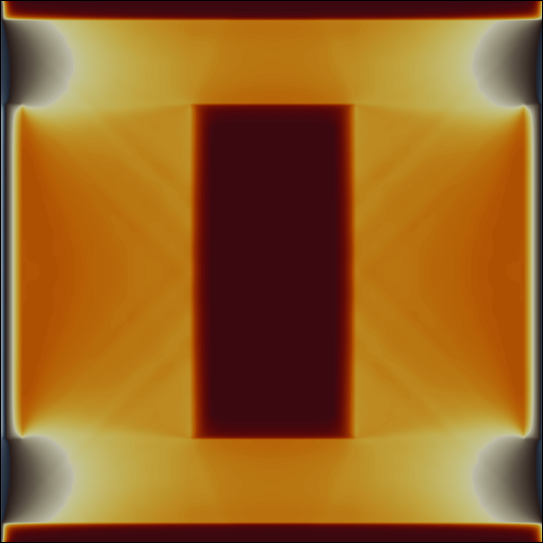}
    \vfill
    \caption{S$_{10}$}
    \label{fig_hohlraum_S_8}
\end{subfigure}
\hfill
\begin{subfigure}[t!]{0.32\linewidth}
  \centering
    \includegraphics[height=5.05cm]{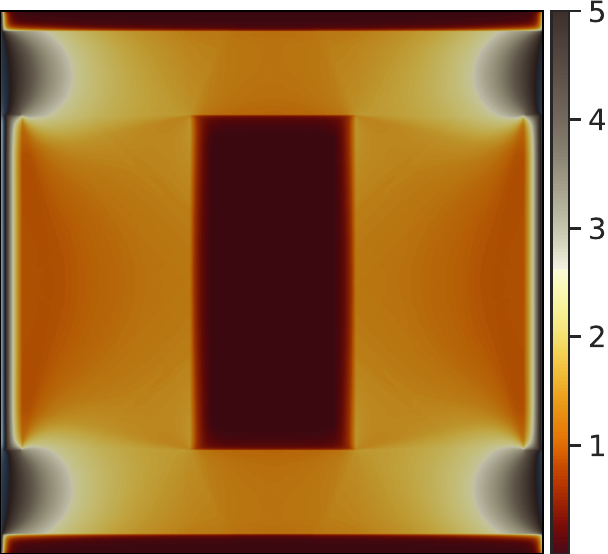}
    \vfill
    \caption{S$_{22}$}
    \label{fig_hohlraum_S_{12}}
\end{subfigure}
\caption{Hohlraum benchmark, second-order finite volume simulation. First row: Different spatial resolutions combined with an $S_6$ angular discretization. Ray effects become apparent as the spatial resolution increases. 
Second row: Different quadrature orders and $h=7.5 \times 10^{-4}$ fixed. Low-order \sn simulations, e.g. $S_2$ exhibit strong ray effects.  These artifacts are particularly pronounced in low-scattering regimes. Thus, high order \sn simulations are required to produce reasonable reference solutions.}\label{fig_example_simulations_hohlraum}
\end{figure}

One of the most important considerations in performing the simulations is how to allocate resources.  Indeed, while the angular and spatial resolution can be varied independently, the optimal choice for a given budget is often not obvious.  One reason for this difficulty is the presence of ray effects---numerical artifacts caused by the imprinting of the angular ordinates onto the solution when there in insufficient resolution \cite{lathrop1971remedies}.  These artifacts are induced by localized sources and discontinuous material cross-sections, and they are most prevalent when the amount of scattering, which has a smoothing effect, is small.  At the same time, spatial discretization errors for the finite volume scheme also have a smoothing effect that can wash out ray effects.  As a result, increased spatial resolution can sometimes increase localized errors; see e.g., the first row of \Cref{fig_example_simulations_hohlraum}.

Our approach to refining the spatial and angular resolution is somewhat adhoc.  While greedy approaches exist, they often rely on the assumption of monotonicity in the error.  Moreover, the metric for defining greedy choices depends on the quantity of interest, of which there are several.  Roughly speaking, we choose a spatial resolution $h$ and then refine the angular quadrature until the successive differences in all the QoIs are small, at which point we refine the spatial mesh again and repeat until the available resources are exhausted.  A demonstration of the angular refinement for the Hohlraum benchmark is provided in the second row of \Cref{fig_example_simulations_hohlraum}.  We use this approach to identify a $3 \times 3$ block of space-angle indexes that contains what we consider to be ``highly resolved references solutions".  We use this block to define a variance for the each QoI, which is used as a proxy for the error in numerical solution.  The blocks, and the QoIs associated with them, are shown in \Cref{fig_lattice_qoi_grid} and \Cref{fig_hohlraum_qoi_grid} for the Lattice and Hohlraum benchmarks, respectively.

\subsection{Lattice benchmark}

\begin{figure}
    \centering
    \begin{subfigure}{0.40\textwidth}
            \includegraphics[width=\textwidth]{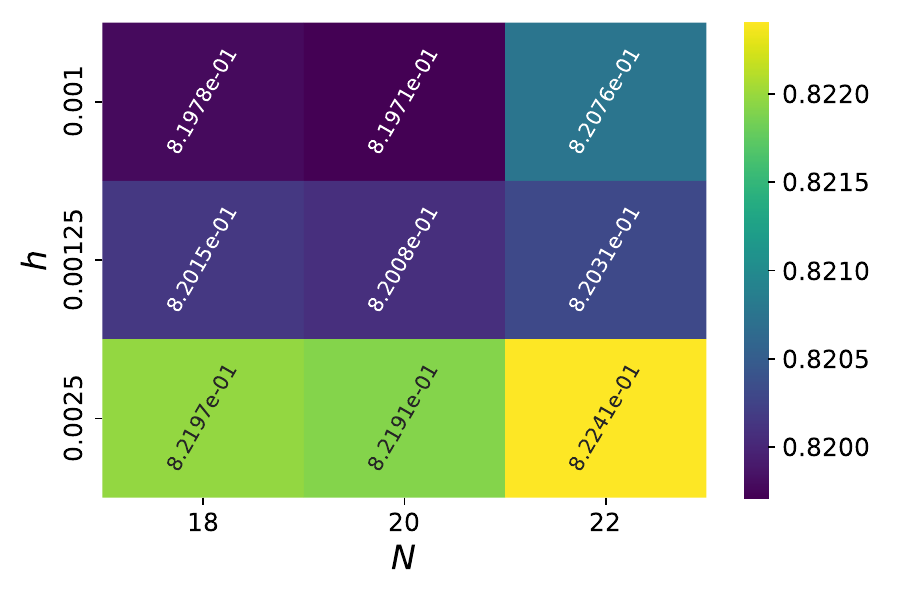}   \caption{QoI-1, $\ell=1.5$ }
    \end{subfigure}
    \hspace{30pt}
    \begin{subfigure}{0.40\textwidth}
            \includegraphics[width=\textwidth]{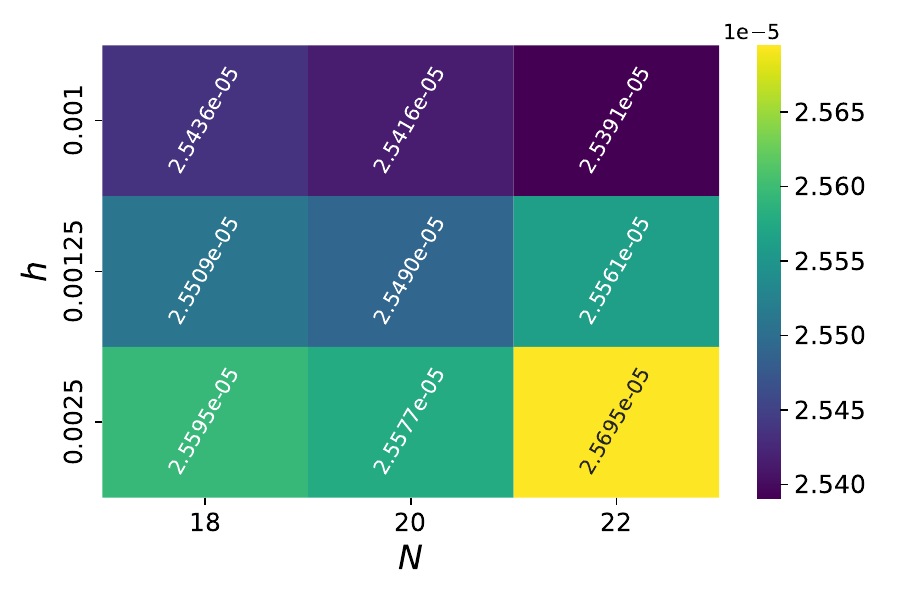}   \caption{QoI-1, $\ell=2.5$   }
    \end{subfigure}
    
    \begin{subfigure}{0.40\textwidth}
            \includegraphics[width=\textwidth]{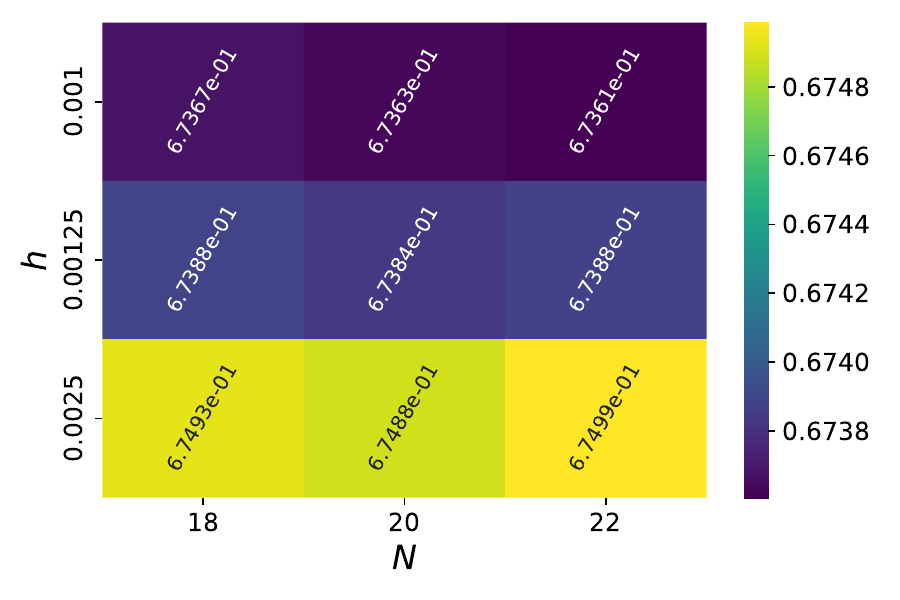}     \caption{QoI-2, $\ell=1.5$  }
    \end{subfigure}
    \hspace{30pt}
        \begin{subfigure}{0.40\textwidth}
            \includegraphics[width=\textwidth]{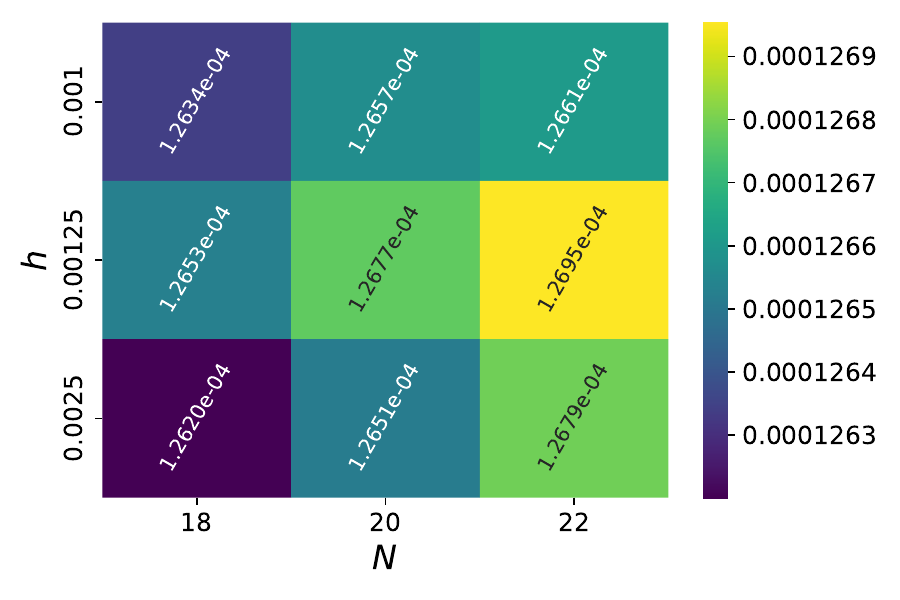}    \caption{QoI-2, $\ell=2.5$ }
    \end{subfigure}
    
        \begin{subfigure}{0.40\textwidth}
            \includegraphics[width=\textwidth]{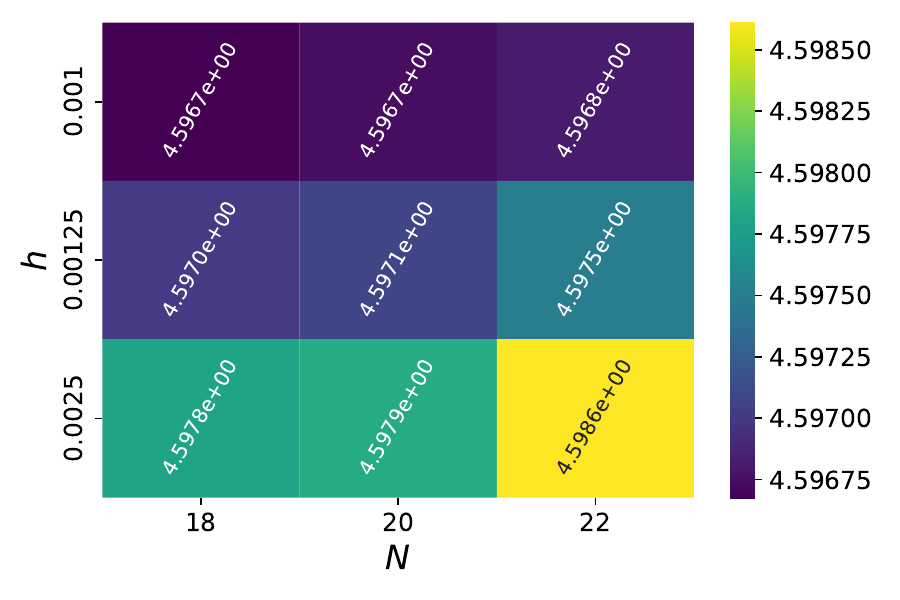}      \caption{QoI-3}
    \end{subfigure}
    \hspace{30pt}
    \begin{subfigure}{0.40\textwidth}
            \includegraphics[width=\textwidth]{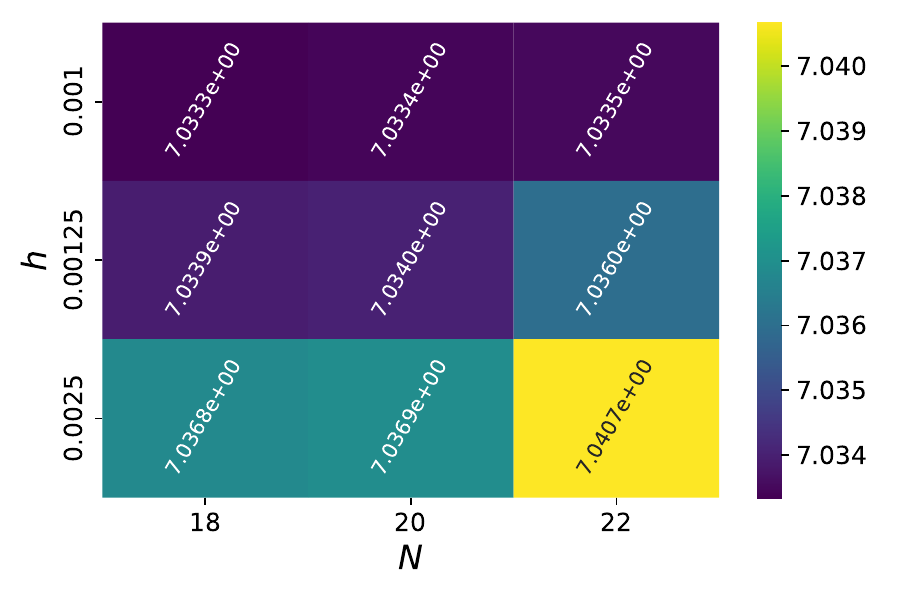}                              \caption{QoI-4}
    \end{subfigure}
        \begin{subfigure}{0.40\textwidth}
            \includegraphics[width=\textwidth]{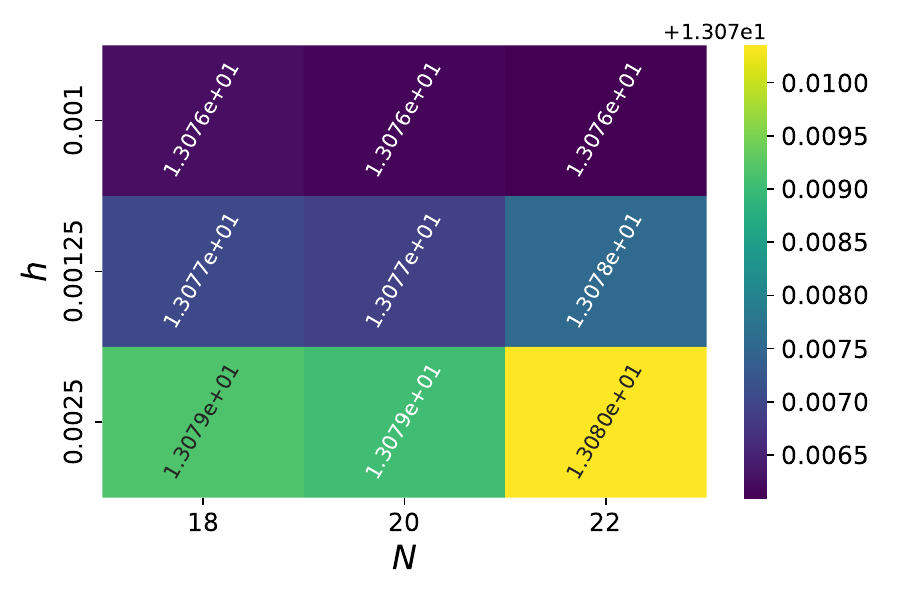}                  \caption{QoI-5}
    \end{subfigure}
 \caption{Quantities of interest for the Lattice benchmark.  Results are computed using a tessellation \sn quadrature for the angular discretization and a finite volume scheme that is a second-order in space and time. Values for each of nine different space-angle resolutions are used to compute mean values and standard deviations, which are summarized in \Cref{tab_summary_lattice}.}
    \label{fig_lattice_qoi_grid}
\end{figure}
We report the quantities of interest as the mean of the nine reference simulations. The reference simulations use the nine high-resolution phase space discretizations with quadrature orders $N \in \{18,20,22\}$ and mesh sizes $h \in \{2.5\times 10^{-3},1.25\times 10^{-3},1\times 10^{-3}\}$.  The finest resolution ($N=22$ and $h=1\times 10^{-3}$) uses a total of  $3.272\times 10^{10}$ degrees of freedom to discretize the phase space.  

\Cref{fig_lattice_qoi_grid} shows the values for each QoI from \Cref{subsubsec_hohlraum_QOI} at each of the nine different reference resolutions.
The mean and standard deviation for each QoI is displayed in \Cref{tab_summary_lattice}.  In each case, the standard deviation over the nine solutions is less than 1\% of the mean, and often it is much less.

\begin{table}[]
    \centering
\include{assets/tables/tab_lattice}
\caption{Lattice benchmark, summary of simulation results. Reported are the mean and standard deviation of each quantity of interest (QoI) of the nine reference solutions. The standard deviation across the nine simulations is typically three orders of magnitude or more smaller than the mean value.}
    \label{tab_summary_lattice}
\end{table}

\clearpage

\subsection{Hohlraum benchmark}

We report the quantities of interest as the mean of nine reference simulations. The reference simulations use the nine discretizations in space and angle given by the quadrature orders $N \in \{18,20,22\}$ and mesh cell  sizes  $h \in \{1\times 10^{-3},8\times 10^{-4},7.5\times 10^{-4}\}$.  The finest resolution ($N=22$ and $h=7.5\times 10^{-4}$) uses a total of  $1.01\times 10^{10}$ degrees of freedom to discretize the phase space.

\begin{figure}
    \centering
    \begin{subfigure}{0.32\textwidth}
            \includegraphics[width=\textwidth]{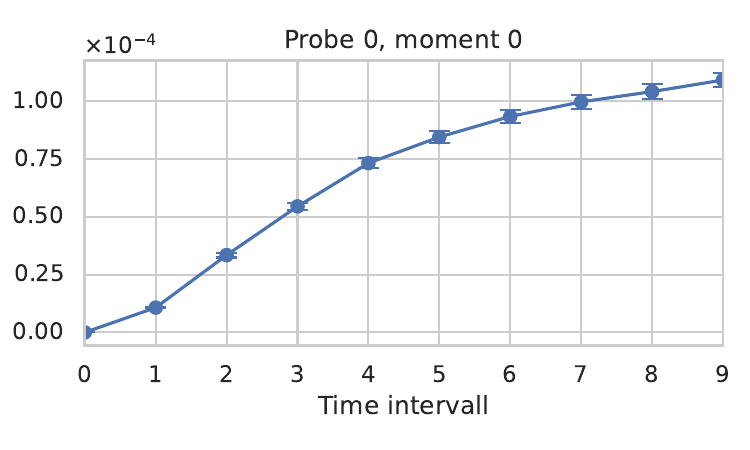}
    \end{subfigure}
    \begin{subfigure}{0.32\textwidth}
            \includegraphics[width=\textwidth]{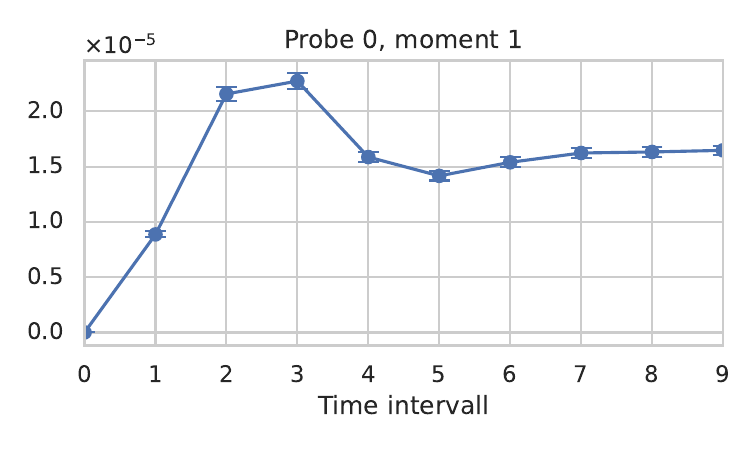}
    \end{subfigure}
    \begin{subfigure}{0.32\textwidth}
            \includegraphics[width=\textwidth]{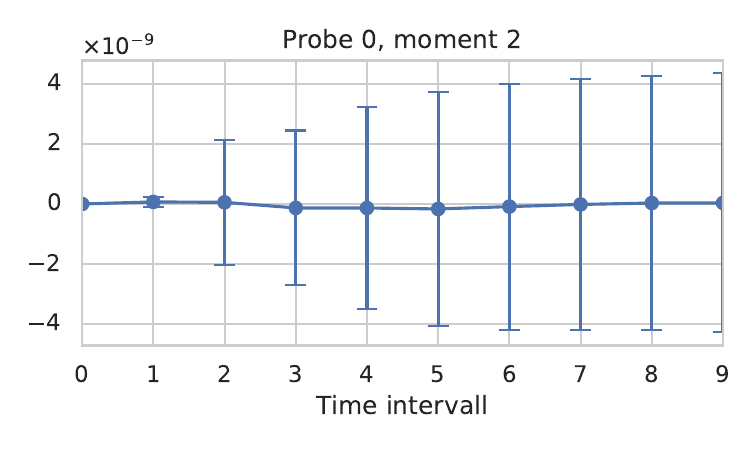}
    \end{subfigure}
    
    \begin{subfigure}{0.32\textwidth}
            \includegraphics[width=\textwidth]{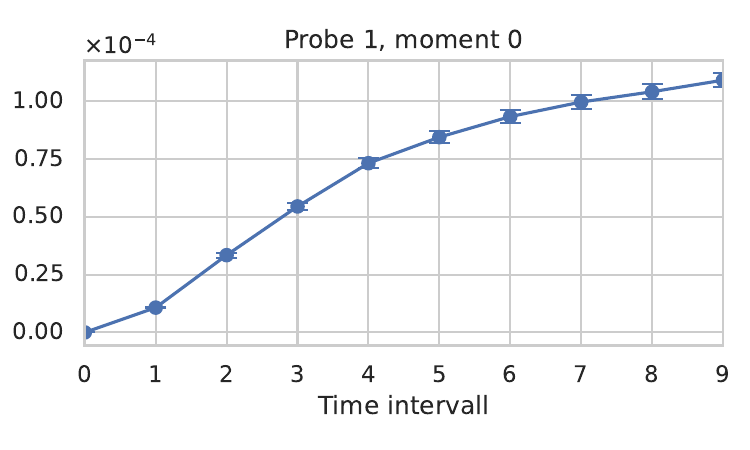}
    \end{subfigure}
        \begin{subfigure}{0.32\textwidth}
            \includegraphics[width=\textwidth]{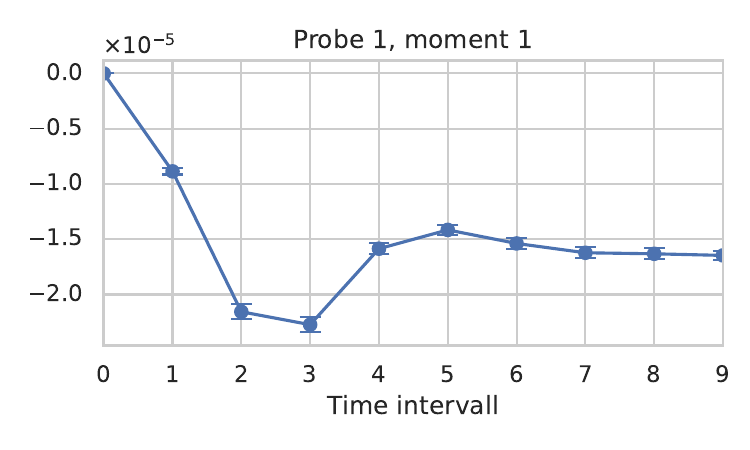}
    \end{subfigure}
        \begin{subfigure}{0.32\textwidth}
            \includegraphics[width=\textwidth]{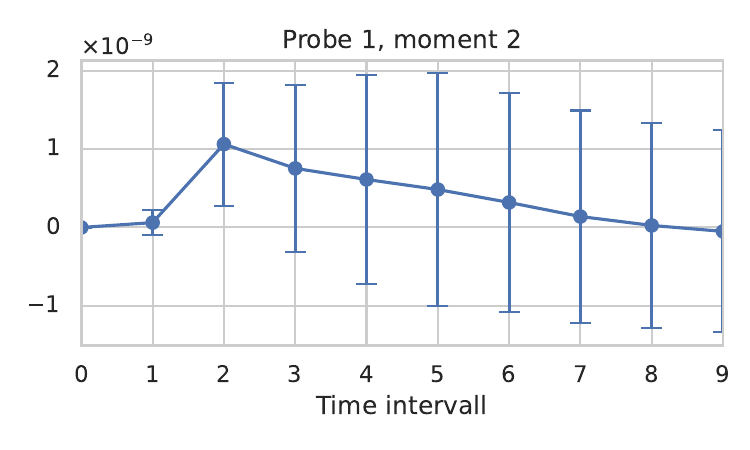}
    \end{subfigure}

     \begin{subfigure}{0.32\textwidth}
            \includegraphics[width=\textwidth]{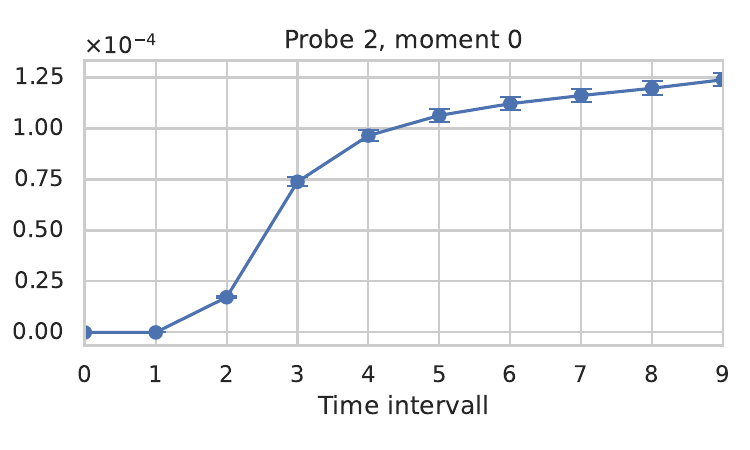}
    \end{subfigure}
    \begin{subfigure}{0.32\textwidth}
            \includegraphics[width=\textwidth]{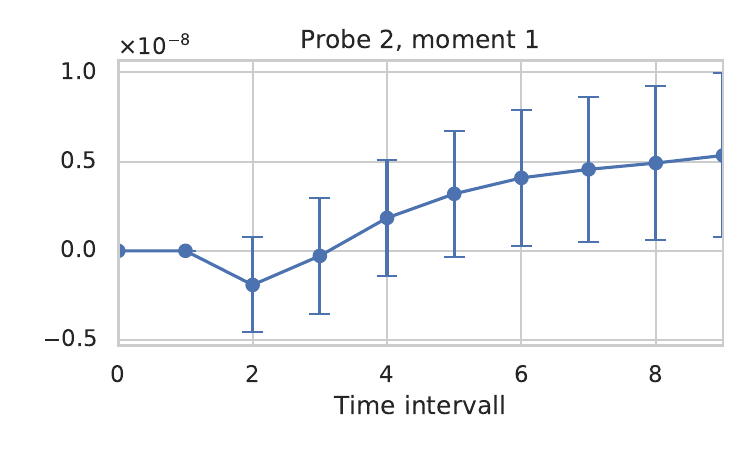}
    \end{subfigure}
    \begin{subfigure}{0.32\textwidth}
            \includegraphics[width=\textwidth]{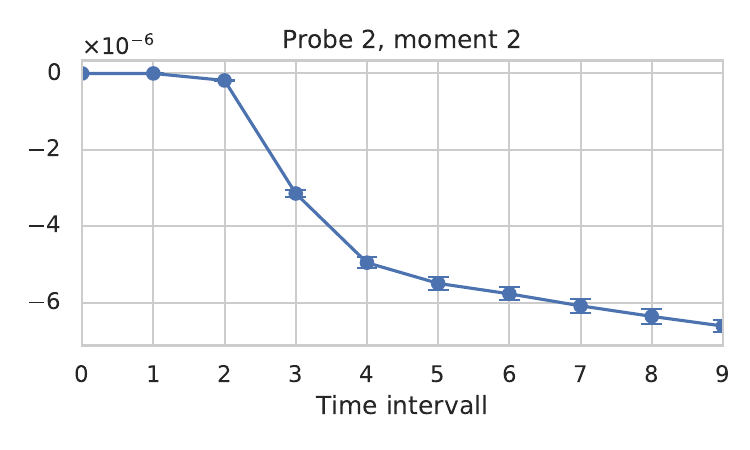}
    \end{subfigure}
    
    \begin{subfigure}{0.32\textwidth}
            \includegraphics[width=\textwidth]{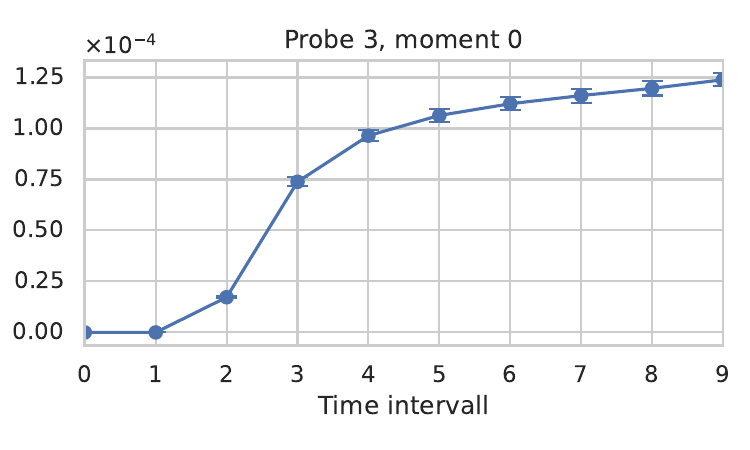}
    \end{subfigure}
        \begin{subfigure}{0.32\textwidth}
            \includegraphics[width=\textwidth]{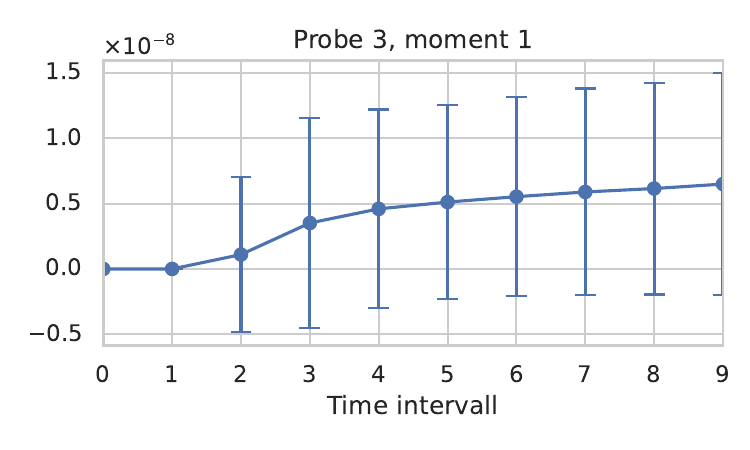}
    \end{subfigure}
        \begin{subfigure}{0.32\textwidth}
            \includegraphics[width=\textwidth]{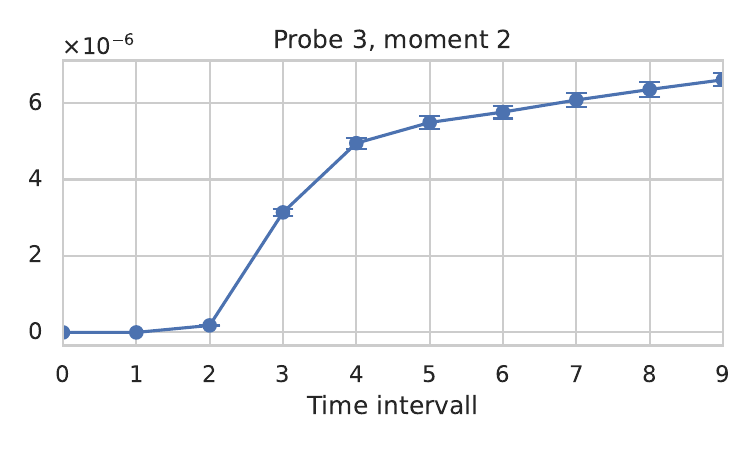}
    \end{subfigure}
    \caption{Time traces of the first three moments with standard deviation at the four probes with coordinates  $(x_1,y_1) = (-0.4,0)$, $(x_2,y_2)  = (0.4,0) $, $(x_3,y_3) = (0,-0.5)$, $(x_4,y_4)  = (0, 0.5)$ in the Hohlraum benchmark simulated with a second-order finite volume \sn scheme.
    The variance of the reported reference values is below $1\times 10^{-5}$ for all probe moments. 
    The symmetry of the benchmark is captured by the fact that the first and third moment of probe 0 and probe 1 are identical, and the second moment switches signs. Similarly, the first two moments of probe 2 and 3 are identical and the third moment switches sign. }
    \label{fig_hohlraum_time_traces}
\end{figure}

The first three moments at the prescribed probe points for the time $10$ time intervals, i.e. QoI-1,  are displayed in \Cref{fig_hohlraum_time_traces}. One can observe the intrinsic symmetry of the Hohlraum benchmark since the first and third moment of the first probe equals the first and third moment of the second probe and the second moment switches sign. Further observe that the third moment is close to zero, since the moments in $y$-direction are eliminated. Similar observations can be made for the third and fourth probe, where the moments in $x$-direction are eliminated, the first moments are equal, and the sign switches at the third moment.

\begin{figure}
    \centering
    \begin{subfigure}{0.40\textwidth}
            \includegraphics[width=\textwidth]{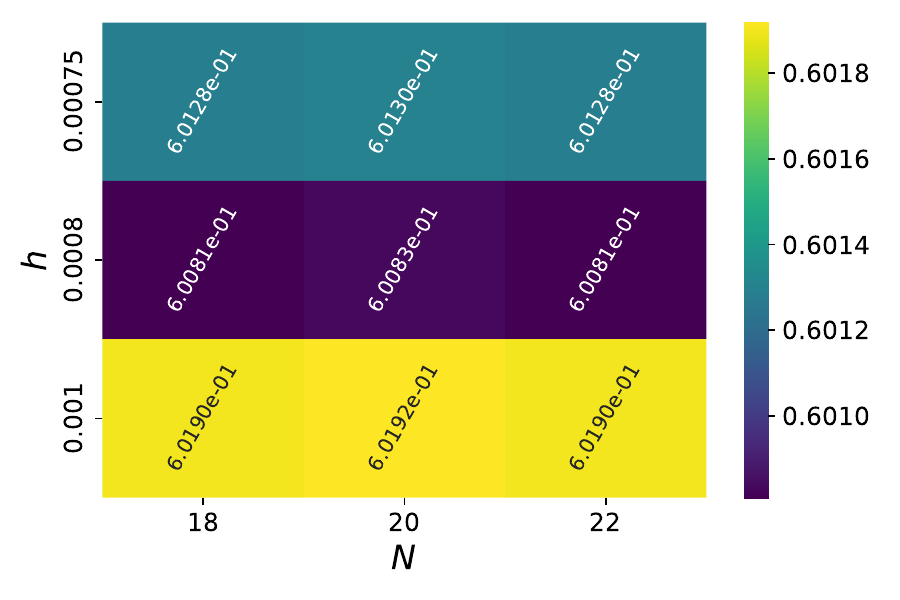}
            \caption{QoI-2, $G\cup B$}
    \end{subfigure}
    \hspace{30pt}
    \begin{subfigure}{0.40\textwidth}
            \includegraphics[width=\textwidth]{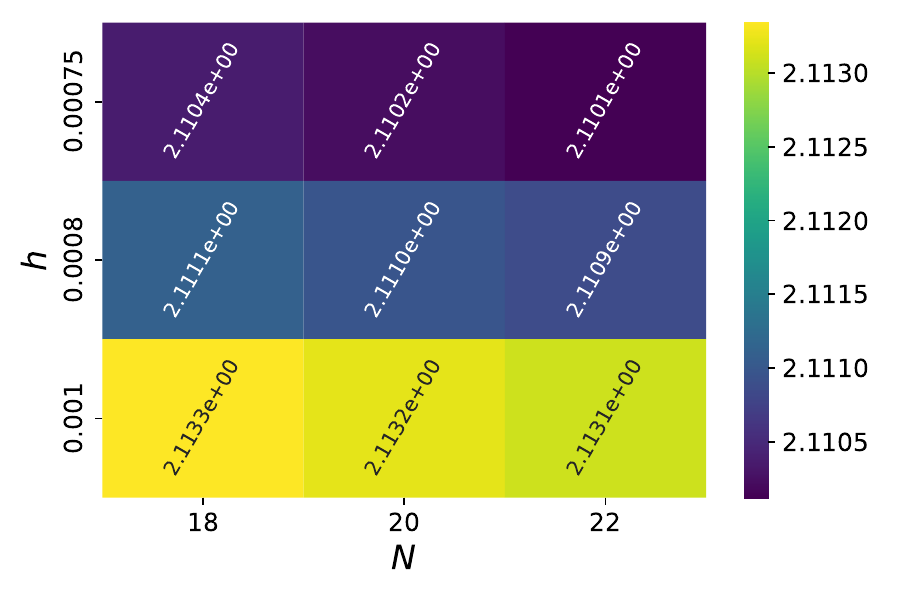}
            \caption{QoI-2, $K$ }
    \end{subfigure}
    
    \begin{subfigure}{0.40\textwidth}
            \includegraphics[width=\textwidth]{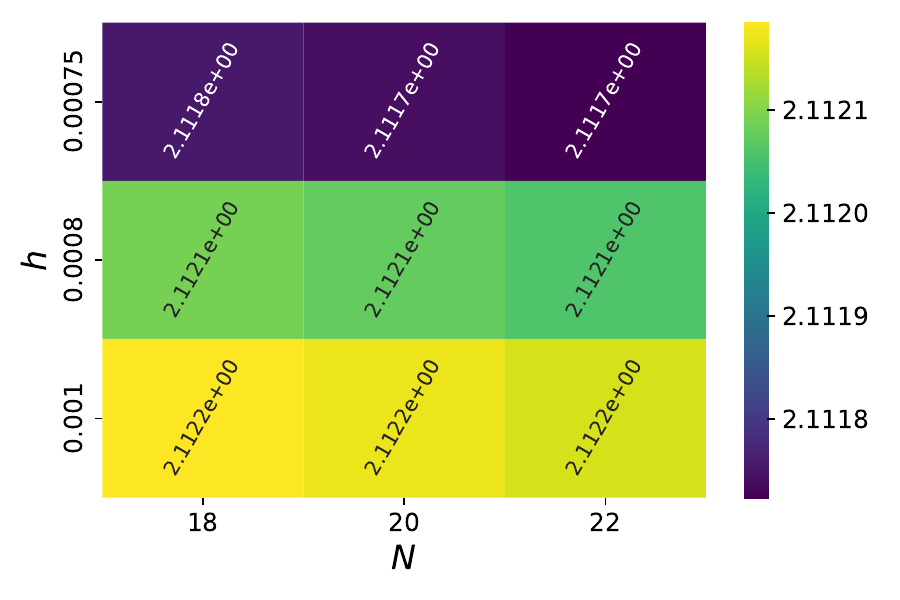}
            \caption{QoI-2, $R$ }
    \end{subfigure}
    \hspace{30pt}
        \begin{subfigure}{0.40\textwidth}
            \includegraphics[width=\textwidth]{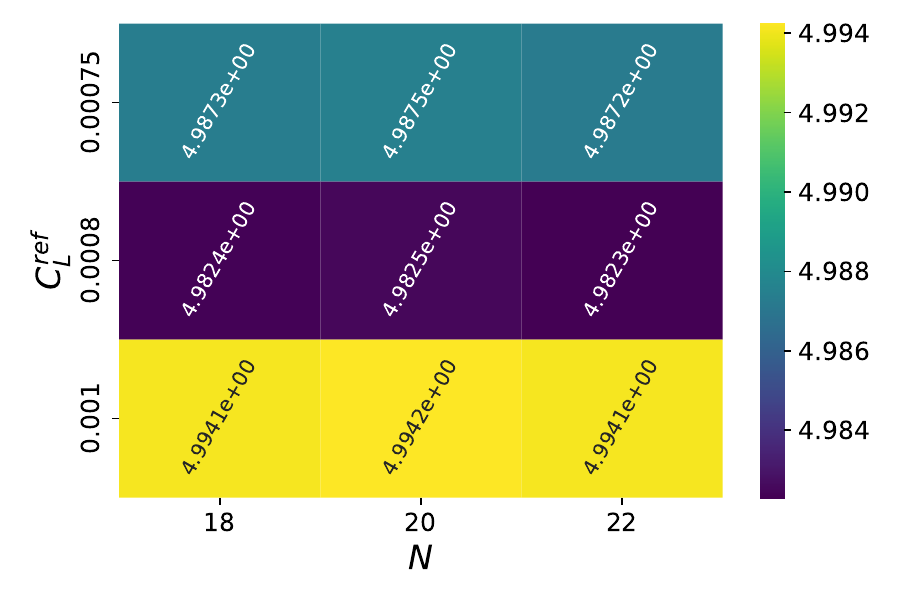}
             \caption{QoI-3 }
    \end{subfigure}
 
        \begin{subfigure}{0.40\textwidth}
            \includegraphics[width=\textwidth]{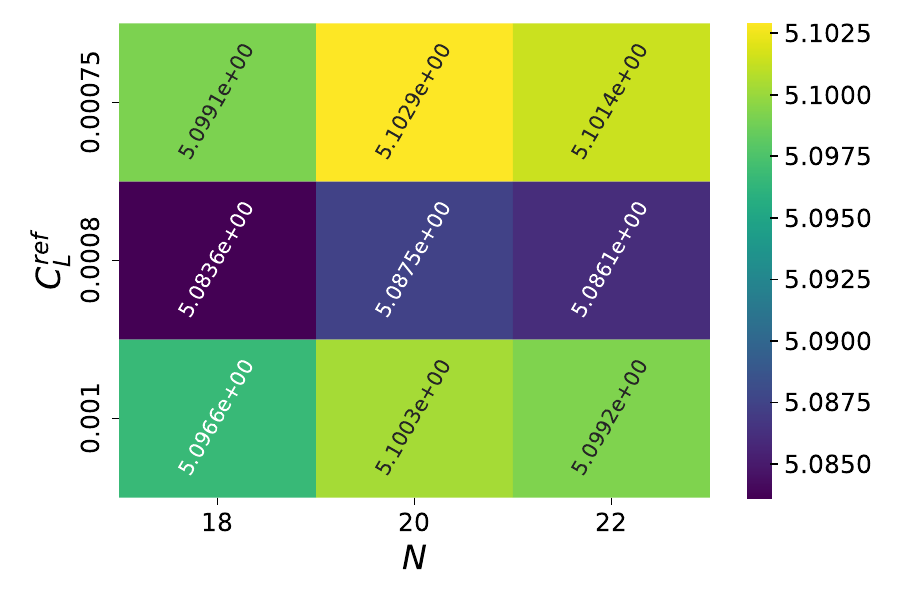}
              \caption{QoI-4}
    \end{subfigure}
    \hspace{30pt}
    \begin{subfigure}{0.40\textwidth}
            \includegraphics[width=\textwidth]{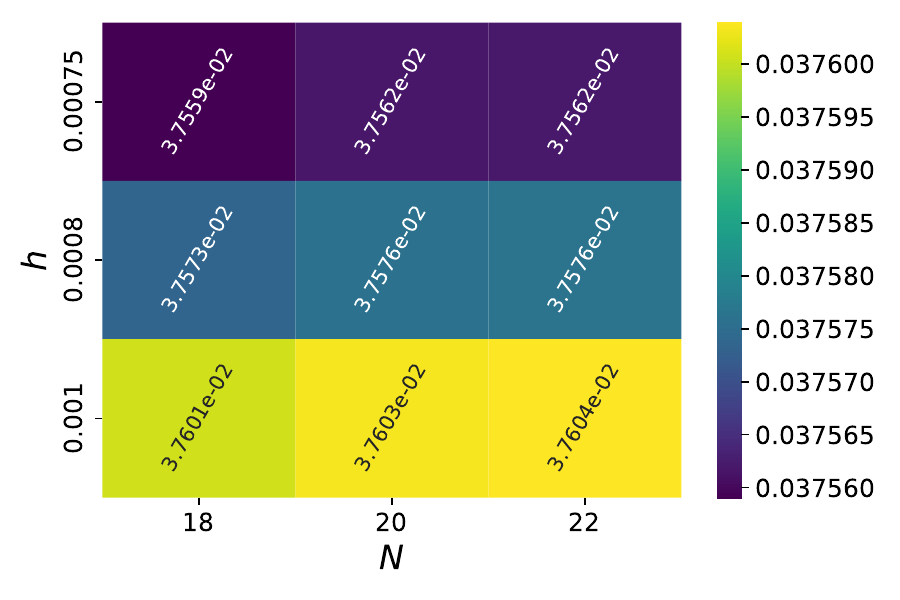}
                                      \caption{QoI-5}
    \end{subfigure}
    
        \begin{subfigure}{0.40\textwidth}
            \includegraphics[width=\textwidth]{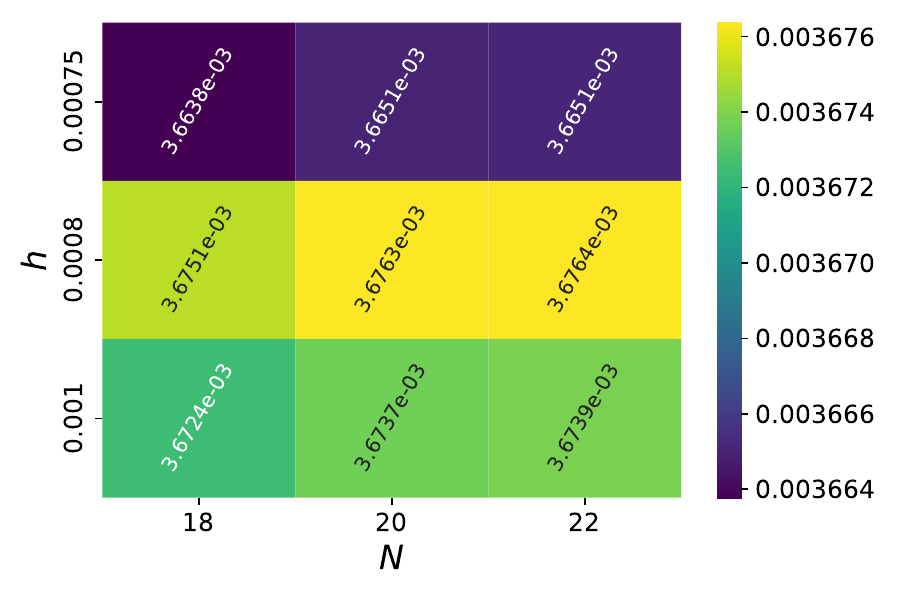}
                          \caption{QoI-6}
    \end{subfigure}
        \hspace{30pt}
        \begin{subfigure}{0.40\textwidth}
            \includegraphics[width=\textwidth]{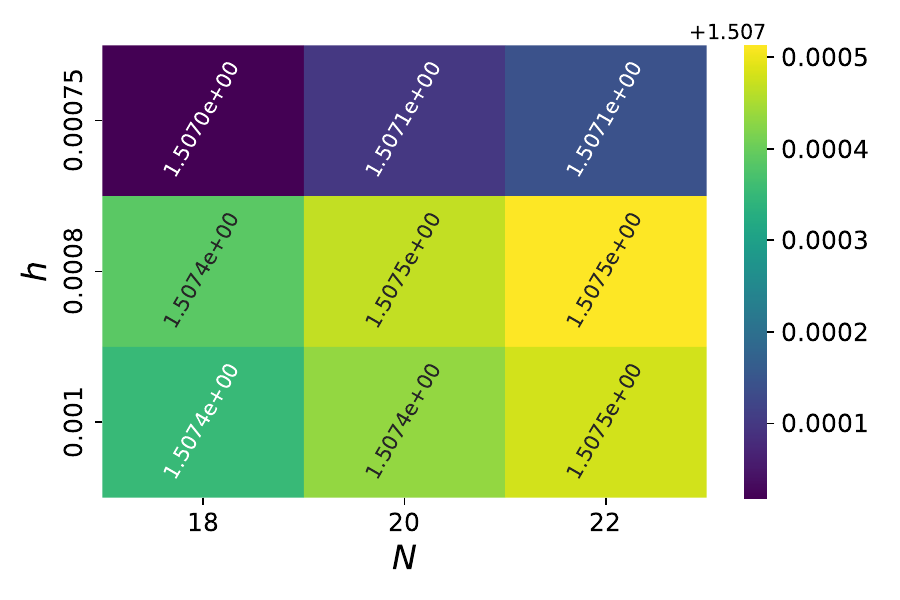}
            \caption{QoI-7}
    \end{subfigure}

    \caption{Quantities of interest for the Hohlraum benchmark.  Results are computed using a tessellation \sn quadrature for the angular discretization and a finite volume scheme that is a second-order in space and time. Values for each of nine different space-angle resolutions are used to compute mean values and standard deviations, which are summarized in Table \Cref{tab_summary_hohlraum}.}
    \label{fig_hohlraum_qoi_grid}
\end{figure}

\begin{figure}
\begin{subfigure}[t!]{0.45\linewidth}
  \centering
    \includegraphics[width=\textwidth]{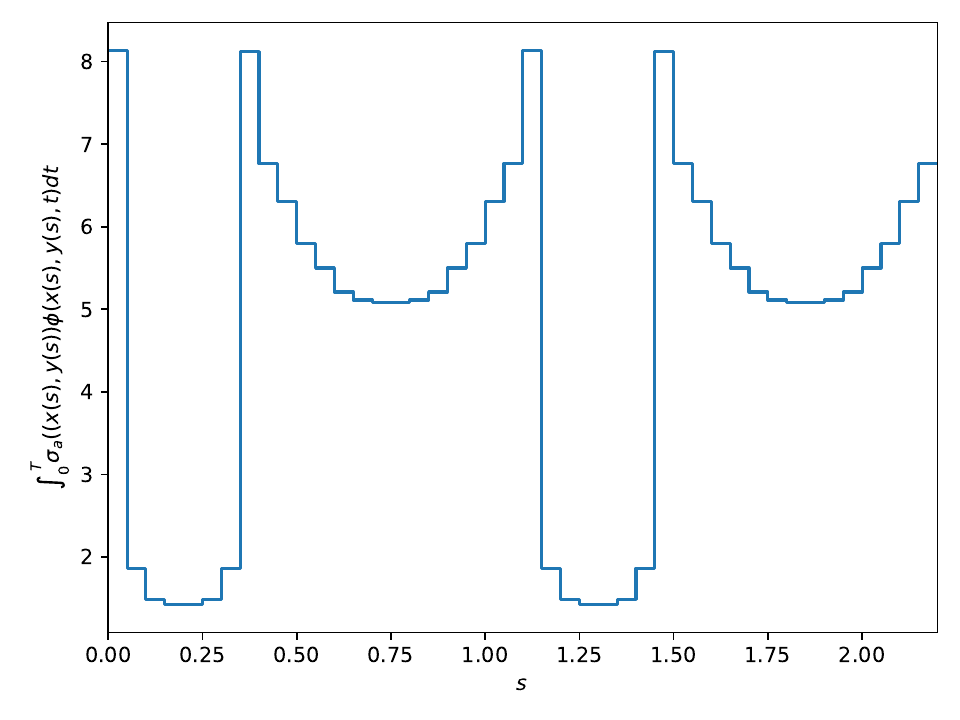}
    \vfill
    \caption{Time integrated absorption values in the green blocks of size $0.05\times 0.05$ for QoI-3.}
   \label{fig_hohlraum_qoi3}
\end{subfigure}
\hspace{30pt}
\begin{subfigure}[t!]{0.45\linewidth}
  \centering
    \includegraphics[width=\textwidth]{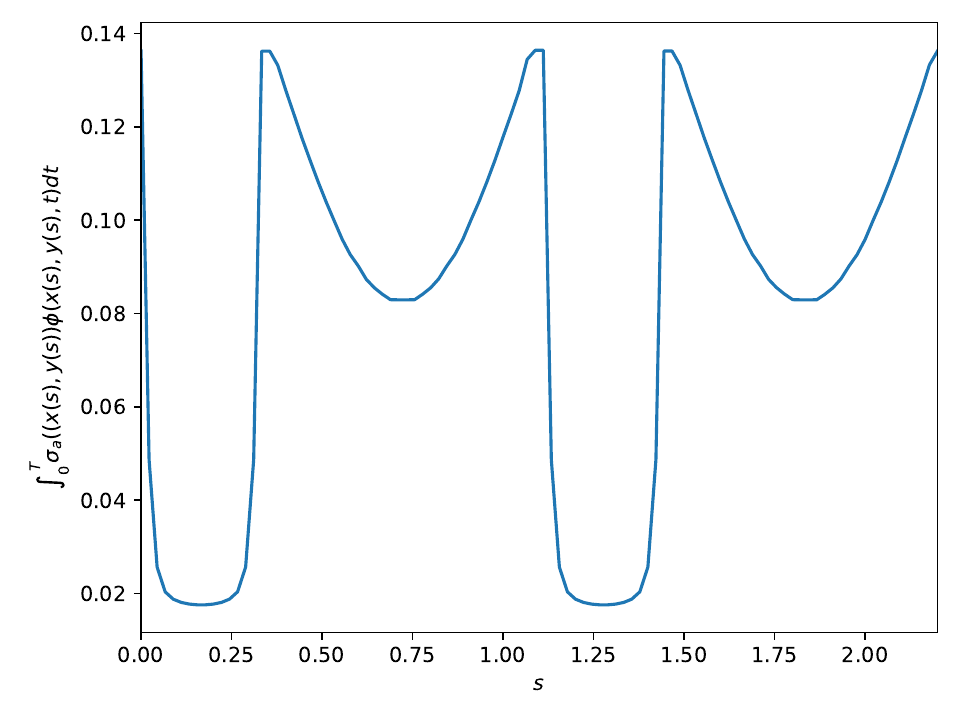}
    \vfill
    \caption{Time integrated absorption values along a line through the center of each green block for QoI-5.}
    \label{fig_hohlraum_qoi5}
\end{subfigure}

\caption{Plots of the time integrated absorption values for QoI-3 (left) and QoI-5 (right) over the location in the green capsule boundary. Each value is the average of the $9$ reference solutions. The spatial position is parameterized by $s\in[0,2.2]$, which describes the path starting from the point $(x,y)=(-0.175,-0.375)$ and traversing the center of the green band counter clockwise. The symmetry of the benchmark causes the periodicity of the plotted absorption values.}\label{fig_line_plots_hohlraum}
\end{figure}

\Cref{fig_hohlraum_qoi_grid} shows the values for each QoI from \Cref{subsubsec_hohlraum_QOI} at each of the nine different reference resolutions.
\Cref{fig_line_plots_hohlraum} shows the time averaged absorption values in the blocks of QoI-3 and the line of QoI-5. The symmetry of the benchmark is evident by the periodicity of the plot.

\begin{table}[]
    \centering
\include{assets/tables/tab_hohlraum}
    \label{tab_summary_hohlraum}
 \caption{ Simulation summary of the Hohlraum benchmark. Reported are the mean and standard deviation of each quantity of interest (QoI) of the nine reference solutions. The standard deviation is consistently is roughly three orders of magnitude or smaller than the mean value.
 For the results of QoI-1; see Figure \ref{fig_hohlraum_time_traces}.}
\end{table}

\section{Conclusions and discussion}
\label{sec_conc}

In this paper, we have constructed high-resolution solutions of the linear radiation transport equation (RTE) for two benchmark problems.  These solutions are used to evaluate various quantities of interest.  The goal of these calculations is to provide near ground-truth comparisons for newer methods current being developed in the community.

Solutions are computed using  a discrete ordinate discretization in angle, a second-order finite volume method on an unstructured grid in space, and a second-order explicit SSP-RK time integrator.  These methods are well-established and are for robustness and stability.   They are implemented using the open-source \texttt{C++}-based solver KiT-RT~\cite{kusch2023kit}, which uses distributed memory to parallelize in angle and shared-memory to parallelize in space.  The code is run on the CADES Baseline cluster of the National Center for Computational Sciences at the Oak Ridge National Laboratory. We use $N_{\rm{MPI}}+1=41$ MPI nodes equipped with 2X AMD 7713 processors with $128$ OpenMP threads.

Data from all the simulations is available at DOI \url{10.13139/OLCF/2540565}.  In addition the QoIs considered in this paper, phase space solutions to the RTE are available at the final run time for each problem.  The code is also available at \url{https:/github.com/KiT-RT/kitrt_code} and \url{https:/github.com/ScSteffen/CharmKiT}.  Thus, if additional resources are available, then even finer solutions can be computed if needed.  The code also allows for changes in the design variables, and hence can also be used as a platform for developing surrogate models, as well as algorithms for ``outer loop" calculations such as uncertainty quantification, design, and eventually control.

Future research efforts will focus on the development of surrogate and algorithms for outer loop calculations.  We will also consider extensions of the simple RTE model to allow for energy groups, different types of radiation, and nonlinear coupling to materials.  

\section{Contribution Statement}
Steffen Schotth\"ofer developed and designed the KiT-RT and CharmKiT simulation code, drafted the manuscript and designed the test-cases.
Cory D. Hauck drafted the manuscript and designed the test-cases. 

\section{Disclosure of Interest}
There are no conflicts of interest to declare.
\section{Declaration of Funding}
This material is based upon work supported by the U.S. Department of Energy, Office of Science, Office of Advanced Scientific Computing Research, as part of their Applied Mathematics Research Program. The work was performed at the Oak Ridge National Laboratory, which is managed by UT-Battelle, LLC under Contract No. De-AC05-00OR22725. The United States Government retains and the publisher, by accepting the article for publication, acknowledges that the United States Government retains a non-exclusive, paid-up, irrevocable, world-wide license to publish or reproduce the published form of this manuscript, or allow others to do so, for the United States Government purposes. The Department of Energy will provide public access to these results of federally sponsored research in accordance with the DOE Public Access Plan (http:/energy.gov/downloads/doe-public-access-plan).
This research used resources of the Compute and Data Environment for Science (CADES) at the Oak Ridge National Laboratory, which is supported by the Office of Science of the U.S. Department of Energy under Contract No. DE-AC05-00OR22725"

\bibliographystyle{plain}
\bibliography{refs}

\end{document}

%% file: macros_steffen.tex

\newcommand{\R}{\mathbb{R}}
\newcommand{\N}{\mathbb{N}}
\newcommand{\C}{\mathbb{C}}

\newcommand{\T}{\mathcal{T}}
\newcommand{\Lagr}{\mathcal{L}}

\newcommand{\dimx}{d} 
\newcommand{\dimp}{n_p}

\newcommand{\ds}{\mathrm{d}s}
\newcommand{\dsig}{\mathrm{d}\sigma}
\newcommand{\dsigD}{\mathrm{d}\sigma_D}
\newcommand{\Stwo}{{\mathbf{S}^2}}

\newcommand{\norm}[1]{\left\lVert#1\right\rVert}
\newcommand{\abs}[1]{\left|#1\right|}
\newcommand{\set}[1]{\left\lbrace#1\right\rbrace}
\newcommand{\limes}[2]{\lim\limits_{{#1}\rightarrow {#2}}}
\newcommand{\intD}{\;\mathrm{d}}
\newcommand{\rom}[1]{\uppercase\expandafter{\romannumeral #1\relax}}
\newcommand{\landauO}{\mathcal{O}}

\newcommand{\inner}[1]{\left< #1 \right>}
\newcommand{\red}[1]{\textcolor{red}{#1}}
\newcommand*\interior[1]{#1^{\mathsf{o}}}
\newcommand{\fu}{\mathbf{u}}
\newcommand{\ofu}{\overline{{\mathbf{u}}}}
\newcommand{\ofus}{{\overline{\mathbf{u}}_{\#}}}
\newcommand{\fv}{\mathbf{v}}
\newcommand{\fw}{\mathbf{w}}
\newcommand{\fy}{\mathbf{y}}
\newcommand{\fz}{\mathbf{z}}
\newcommand{\fx}{\mathbf{x}}
\newcommand{\fm}{\mathbf{m}}
\newcommand{\fg}{\mathbf{g}}
\newcommand{\fa}{\mathbf{a}}
\newcommand{\fb}{\mathbf{b}}
\newcommand{\innerInt}[1]{\int_{\mathbb{S}^2} #1 \intD\fv}

\newcommand{\balpha}{{\boldsymbol{\alpha}}}
\newcommand{\bbeta}{{\boldsymbol{\beta}}}
\newcommand{\blue}[1]{\textcolor{blue}{#1}}
\newcommand{\green}[1]{\textcolor{green}{#1}}
\newcommand{\cdh}[1]{\textcolor{red}{#1}}

\newtheorem{theorem}{Theorem}
\newtheorem{lemma}{Lemma}
\newtheorem{corollary}{Corollary}
\newtheorem{definition}{Definition}
\newtheorem{assumption}{Assumption}
\newcommand{\lrW}{W_r}
\newcommand{\W}{W}
\newcommand{\augUc}{\widetilde{U}_c}
\newcommand{\augSc}{\widetilde{S}_c}
\newcommand{\augVc}{\widetilde{V}_c}
\newcommand{\augWrc}{\widetilde{W}_{r,c}}
\newcommand{\augU}{{\widetilde{U}}}
\newcommand{\augS}{{\widetilde{S}}}
\newcommand{\hatS}{{\widehat{S}}}
\newcommand{\hatSc}{{\widehat{S}_c}}
\newcommand{\augV}{{\widetilde{V}}}
\newcommand{\augWr}{{\widetilde{W}_{r}}}
\newcommand{\Wr}{{{W}_{r}}}
\newcommand{\augF}{{\widetilde{F}}}

\def\intstep{{\small \textsf{one-step-integrate}}}
\def\timestep{\eta}

\newcommand{\ssnote}[1]{\textcolor{teal}{[{SS}] $\diamondsuit$ #1}}

%% file: assets/tables/tab_lattice.tex
 {
\begin{tabular}{l | ccc }
\toprule   
 
Quantity of interest (QOI)  & Mean & Std. Dev. &  Std. Dev. / Mean  \\
\midrule

QOI-1, $l=1.5$   & $6.737\times 10^{-1}$ & $1.360\times 10^{-4}$ & $2.019\times 10^{-4}$ \\
QOI-2, $l=1.5$   & $8.205\times 10^{-1}$ & $2.275\times 10^{-4}$ & $2.773\times 10^{-4}$ \\
QOI-1, $l=2.5$   & $1.267\times 10^{-4}$ & $1.730\times 10^{-7}$ & $1.365\times 10^{-3}$ \\
QOI-2, $l=2.5$   & $2.547\times 10^{-5}$ & $8.506\times 10^{-8}$ & $3.340\times 10^{-3}$ \\
QOI-3           & $4.597\times 10^{0}$  & $3.354\times 10^{-4}$ & $7.296\times 10^{-5}$ \\
QOI-4           & $7.034\times 10^{0}$  & $3.864\times 10^{-3}$ & $5.496\times 10^{-4}$ \\
QOI-5           & $1.307\times 10^{1}$  & $7.250\times 10^{-4}$ & $5.548\times 10^{-5}$ \\
\midrule
Quadrature order $N$ &\multicolumn{3}{c}{$18, 20, 22$} \\
Cell size $h$  &\multicolumn{3}{c}{$2.5\times 10^{-3}$, $1.25\times 10^{-3}$, $1\times 10^{-3}$ } \\
Max. degrees of freedom &\multicolumn{3}{c}{$3.272\times 10^{10}$} \\
 \bottomrule
\end{tabular}
}

%% file: assets/tables/tab_hohlraum.tex
\begin{tabular}{l | ccc }
\toprule   
Quantity of interest (QOI)  & Mean & Std. Dev.&  Std. Dev. / Mean\\
\midrule
QOI-2, $G\cup B$  & $6.010 \times 10^{-1}$ & $2.343 \times 10^{-4}$ & $3.900 \times 10^{-4}$ \\
QOI-2, $K$    & $2.110 \times 10^0$ & $1.670 \times 10^{-4}$ & $7.914 \times 10^{-5}$ \\
QOI-2, $R$   & $2.110 \times 10^0$ & $3.669 \times 10^{-4}$ & $1.739 \times 10^{-4}$ \\
QOI-3 & $4.984 \times 10^0$ & $2.352 \times 10^{-3}$ & $4.719 \times 10^{-4}$ \\
QOI-4 & $5.093 \times 10^0$ & $6.625 \times 10^{-3}$ & $1.301 \times 10^{-3}$ \\
QOI-5 & $3.756 \times 10^{-2}$ & $7.150 \times 10^{-6}$ & $1.904 \times 10^{-4}$ \\
QOI-6 & $3.670 \times 10^{-3}$ & $5.618 \times 10^{-6}$ & $1.531 \times 10^{-3}$ \\
QOI-7 & $1.507 \times 10^0 $ & $1.845 \times 10^{-4}$ & $1.224 \times 10^{-4}$ \\

\midrule
Quadrature orders $N$ &\multicolumn{3}{c}{$18,20,22$} \\
Cell size $Ch$ &\multicolumn{3}{c}{$1 \times 10^{-3}, 8 \times 10^{-4}, 7.5 \times 10^{-4}$} \\
Max. degrees of freedom &\multicolumn{3}{c}{$1.378 \times 10^{10}$} \\
\bottomrule
\end{tabular}